\documentclass[11pt,dvips]{article}
\usepackage{rotating}
\usepackage{hhline}
\usepackage{array}
\usepackage{epsfig}
\usepackage{amssymb}

\advance\textheight by \topskip
\begin{document}
\tolerance=100000
\thispagestyle{empty}
\setcounter{page}{0}

\newcommand{\nn}{\nonumber}
\newcommand{\be}{\begin{equation}}
\newcommand{\ee}{\end{equation}}
\newcommand{\ba}{\begin{eqnarray}}
\newcommand{\ea}{\end{eqnarray}}
\newcommand{\bann}{\begin{eqnarray*}}
\newcommand{\eann}{\end{eqnarray*}}
\newcommand{\bc}{\begin{center}}
\newcommand{\ec}{\end{center}}
\newcommand{\ds}{\displaystyle}
\newcommand{\half}{\frac{1}{2}}
\newcommand{\sts}{\scriptstyle}
\newcommand{\ngs}{\!\!\!\!\!\!}
\newcommand{\rb}[2]{\raisebox{#1}[-#1]{#2}}
\newcommand{\CP}{${\cal CP}$~}
\newcommand{\cosb}{\cos \beta}
\newcommand{\sinb}{\sin \beta}
\newcommand{\ml}{\frac{1}{\sqrt{2}} \lambda v}
\newcommand{\sbomu}{\frac{\sin 2 \beta}{2 \mu}}
\newcommand{\kmol}{\frac{\kappa \mu}{\lambda}}
\begin{titlepage}

\begin{flushright}
CERN-TH/2003-077\\
DESY 03-066\\
ITEP-03-05\vspace{0.5cm}\\
\end{flushright}

\vspace{1cm}

\renewcommand{\thefootnote}{\fnsymbol{footnote}}
\begin{center}
  {\Large \bf The Higgs Sector \\ of the Next-to-Minimal Supersymmetric Standard Model}\\[1cm]
{\large D.J.~Miller$^{1}$, R.~Nevzorov$^{2}$ \\ 
and P.M.~Zerwas$^{3}$}\\[1cm]
 {\it 
$^1$ Theory Division, CERN, CH-1211 Geneva 23, Switzerland\\
$^2$ ITEP, Moscow, Russia\\
$^3$ Deutsches Elektronen--Synchrotron DESY, D--22603 Hamburg, 
Germany}\\
\end{center}

\renewcommand{\thefootnote}{\arabic{footnote}}
\vspace{3cm}

\begin{abstract}
\noindent
The Higgs boson spectrum of the Next-to-Minimal Supersymmetric
Standard Model is examined. The model includes a singlet Higgs field
$S$ in addition to the two Higgs doublets of the minimal
extension. `Natural' values of the parameters of the model are
motivated by their renormalization group running and the vacuum
stability.  The qualitative features of the Higgs boson masses are
dependent on how strongly the Peccei-Quinn $U(1)$ symmetry of the
model is broken, measured by the self-coupling of the singlet field in
the superpotential. We explore the Higgs boson masses and their
couplings to gauge bosons for various representative scenarios.

\end{abstract}

\end{titlepage}

\section{Introduction}

Supersymmetric models~\cite{susy,susy_models} take an important step
toward a solution of the hierarchy problem~\cite{hierarchy_problem} by
stabilizing the ratio of the electroweak and Plank/GUT scales. In the
Standard Model (SM), the Higgs boson mass gains radiative corrections
which depend quadratically on the cut-off scale of the theory (usually
taken to be the GUT scale $M_{\rm GUT}\simeq 2 \times 10^{16}$ GeV),
threatening to generate a mass which is far too large to explain
electroweak symmetry breaking. In order to stabilize this mass at a
phenomenologically acceptable scale (below a TeV), one must fine tune
the parameters of the model, creating an accidental cancellation
between the bare mass and the quantum corrections. By introducing a
new symmetry, a supersymmetry~\cite{susy} between bosons and fermions,
one introduces new contributions to the quantum corrections to the
Higgs mass. The supersymmetry ensures that the contribution of new
supersymmetric particles exactly cancel the quadratic divergence of
their Standard Model partners, removing the sensitivity of the Higgs
mass to the cut-off scale in a natural way.

The Minimal Supersymmetric Standard Model (MSSM) contains a scale
$\mu$, the Higgs-higgsino mass parameter in the superpotential, which
is phenomenologically constrained to lie not far from the electroweak
scale~\cite{mu_problem}; in the Next-to-Minimal Supersymmetric
Standard Model (NMSSM) this mass parameter may be linked dynamically
to the electroweak scale in a natural
way~\cite{nmssm1}--\cite{nmssm2}.  The superpotential of
the MSSM must be analytic in the fields, preventing the use of only
one Higgs doublet for the generation of both up-type and down-type
quark masses. Thus the model requires two Higgs doublets, which is
also necessary to maintain an anomaly free theory.  One of the
doublets ($H_u$) provides a mass for up-type quarks while the other
($H_d$) provides a mass for down-type quarks and charged leptons.  The
term $\mu H_u H_d$ in the superpotential of the MSSM mixes the two
Higgs doublets. Since the parameter $\mu$, present before the symmetry
is broken, has the dimension of mass, one would naturally expect it to
be either zero or the Planck scale ($M_{\rm Pl}$). However, if $\mu=0$
then the form of the renormalization group equations~\cite{falck}
implies that the mixing between Higgs doublets is not generated at any
scale; the minimum of the Higgs potential occurs for $\langle H_d
\rangle=0$, causing the down-type quarks and charged leptons to remain
massless after symmetry breaking. In the opposite case, for $\mu\simeq
M_{\rm Pl}$, the Higgs scalars acquire a huge contribution $\mu^2$ to
their squared masses and the fine tuning problem is
reintroduced. Indeed, one finds that $\mu$ is required to be of the
order of the electroweak scale in order to provide the correct pattern
of electroweak symmetry breaking.

The most elegant solution\footnote{For other solutions to the
$\mu$-problem see Ref.\cite{other_solutions}.} of this $\mu$--problem
is to introduce a new singlet Higgs field $S$, and replace the
$\mu$-term $\mu H_u H_d$ by an interaction $\lambda S (H_u H_d)$. When
the extra scalar field $S$ acquires a non-zero vacuum expectation
value (VEV) an effective $\mu$-term of the required size is
automatically generated, and the effective $\mu$ parameter may then
naturally be expected to be of the electroweak scale, $\mu=\lambda
\langle S \rangle$. In this way $\mu$ can be linked dynamically to the
scale of electroweak symmetry breaking (although the mechanism
generating the common scale of order $100$~GeV to $1$~TeV is left
unexplained).  This situation naturally appears in the framework of
superstring--inspired $E_6$ models~\cite{E6}. At the string scale, the
$E_6$ symmetry can be broken, leading to the usual low energy gauge
groups with additional $U(1)$ factors (for instance
\mbox{$SO(10) \times U(1) \times \ldots$}). These extra $U(1)$ 
symmetries can be broken via the Higgs mechanism in which the scalar
components of chiral supermultiplets acquire non-zero VEVs in such a
way that all additional gauge bosons and exotic particles acquire huge
masses and decouple, except for one singlet field $S$ and a pair of
Higgs doublets $H_u$ and $H_d$.

However, this model still possesses an extra global $U(1)$ symmetry, a
Peccei--Quinn (PQ) symmetry~\cite{PQ_symmetry}, which is explicitly
broken in the MSSM by the $\mu$-term itself. The breaking of this
extra $U(1)$ at the electroweak scale leads to the appearance of a
massless\footnote{The physical axion would acquire a small mass by
mixing with the pion.} CP--odd scalar in the Higgs boson spectrum, the
PQ axion~\cite{PQ_axion}. Unfortunately, this axion leads to
astrophysical and cosmological constraints which rule out most of the
allowed parameter space, leaving only a small window with
$10^{-7}<\lambda<10^{-10}$~\cite{axion_constraints}. Due to the very
small value of $\lambda$, a very large value of $\langle S \rangle$
would be required to generate $\mu$ in the required energy range, and
this model is unsatisfactory as a solution to the $\mu$-problem.
However, this massless axion may be avoided by introducing a term
cubic in the new singlet superfield in the superpotential. This term
explicitly breaks the additional $U(1)$ global symmetry, providing a
mass to the CP--odd scalar. This model is known as the
Next--to--Minimal Supersymmetric Standard Model (NMSSM), and is
described by the superpotential
\be 
W=\hat{u}^c \, \mathbf{h_u} \hat{Q} \hat{H}_u 
-\hat{d}^c \, \mathbf{h_d} \hat{Q} \hat{H}_d 
-\hat{e}^c \, \mathbf{h_e} \hat{L} \hat{H}_d 
+\lambda \hat{S}(\hat{H}_u \hat{H}_d)+\frac{1}{3}\kappa\hat{S}^3. 
\label{eq:superpotential}
\ee

Despite the removal of the PQ symmetry, the NMSSM superpotential is
still invariant under a discrete $\mathbb{Z}_3$ symmetry: $\Phi \to
e^{2 \pi i/3}\Phi$, where $\Phi$ denotes the observable superfields.
Left untamed, this $\mathbb{Z}_3$ symmetry would lead to the formation
of domain walls in the early universe between regions which were
causally disconnected during the period of electroweak symmetry
breaking~\cite{domain_walls}.  Such domain wall structures of the
vacuum create unacceptably large anisotropies in the cosmic microwave
background~\cite{CMB}.  In an attempt to break the $\mathbb{Z}_3$
symmetry, operators suppressed by powers of the Planck scale can be
introduced. However, it has been shown that these operators, in
general, give rise to quadratically divergent tadpole contributions,
which once again lead to a destabilization of the mass
hierarchy~\cite{tadpoles}.  This problem can be circumvented by
introducing new discrete symmetries to forbid or loop suppresses the
dangerous tadpole contributions. In this case the breaking of the
$\mathbb{Z}_3$ symmetry should be small enough to not upset the mass
hierarchy but large enough to prevent the problematic domain
walls~\cite{kill_tadpoles1,kill_tadpoles2}. Indeed, in these scenarios one would
expect the surviving tadpole terms to be sufficiently suppressed to
not effect the low-energy phenomenology described in this study.

We analyze the mass spectrum of the NMSSM paying attention in this
report not only to the lowest mass state but also to the heavier Higgs
bosons. Characteristic mass patterns emerge for natural choices of the
parameters. An important parameter is the extent to which the PQ
symmetry is broken, measured by the size of the dimensionless coupling
$\kappa$. Moreover, the couplings of the Higgs bosons to the $Z$ boson
are studied for several representative scenarios appropriate to the
production of the NMSSM Higgs bosons at the next generation of proton
and electron-positron colliders. 

A central point of the paper is the analytical analyses of the Higgs
mass spectrum and couplings in the approximation where the
supersymmetry scale is large and the Higgs parameter $\tan \beta$ is
moderate to large. The concise formulae which can be derived in this
parameter range describe the system to surprisingly high accuracy
while providing a valuable analytical understanding of the NMSSM Higgs
sector, which is significantly more complicated than the minimal
extension.

The paper is organized as follows. In Sec.(2) we describe the Higgs
sector of supersymmetric models with an additional singlet
superfield. The vacuum is examined and used to place constraints on
the parameter space. The Higgs mass matrices are discussed, and an
approximate diagonalization of the CP--even mass matrix is used to
provide analytic expressions for the masses and mixings, from which
the couplings to the $Z$ boson can be derived.  These expressions are
used to give a qualitative understanding of how the masses and
couplings depend upon the parameters.  In Sec.(3) the Higgs boson
spectrum of the NMSSM is considered for three representative parameter
choices in detail, characterized by the breaking of the PQ
symmetry. Natural values for the parameters are motivated and it is
shown that the renormalization group flow from the GUT scale down to
the electroweak scale prefers scenarios where the PQ symmetry is only
slightly broken. In this case, two CP--even Higgs bosons and the
lightest CP--odd Higgs boson have masses of order $100$~GeV, while one
CP--even, one CP--odd and the charged Higgs bosons are heavy.  The
couplings of the lightest Higgs bosons to the $Z$ are reduced
preventing their detection at LEP if kinematically possible otherwise.
However, their observation is possible at a future linear collider
operating in the TeV energy range with high luminosity, so that the
NMSSM may be distinguished from the MSSM at such a facility.  If the
PQ symmetry is strongly broken for large values of $\kappa$, the extra
fields may become more massive and eventually decouple, making such a
comparison more difficult.  The results are summarized in the
Sec.(4). In the appendix, approximate solutions for the masses and
couplings of CP--even Higgs bosons are deduced for physically
appealing parameter domains.

\section{The NMSSM Higgs Sector}

\subsection{The NMSSM Higgs Potential}

The NMSSM Higgs sector is distinguished from that for the MSSM by the
addition of an extra complex scalar field, $S$. The Higgs fields of
the model then consist of the usual two Higgs doublets together with
this extra Higgs singlet,
\be  
H_u= \left( \begin{array}{c} H_u^+ \\ H_u^0 \end{array} \right), 
\quad H_d= \left( \begin{array}{c} H_d^0 \\ H_d^- \end{array} \right),
\quad S. 
\ee
In the superpotential, already presented in
Eqn.(\ref{eq:superpotential}), the extra singlet is allowed to couple
only to the Higgs doublets of the model, and consequently the
couplings of the new fields to gauge bosons will only be manifest via
their mixing with the other Higgs fields\footnote{Higgs self
interactions and couplings to the sleptons, squarks and higgsinos will
also gain extra contributions in the Lagrangian directly from the new
terms in the Superpotential.}. The superpotential leads to the
tree--level Higgs potential~\cite{ellis}:
\be V = V_F+V_D+V_{\rm soft}, \label{eq:Hpot} \ee
with
\ba
V_F&=& |\lambda S|^2 (|H_u|^2+|H_d|^2) + |\lambda H_uH_d
+\kappa S^2|^2 , \label{eq:HpotF} \\
V_D &=& \frac{1}{8} \bar g^2 (|H_d|^2-|H_u|^2)^2
 +\frac{1}{2}g^2|H_u^{\dagger}H_d|^2 , \label{eq:HpotD} \\
V_{\rm soft}&=&m_{H_u}^2|H_u|^2 + m_{H_d}^2|H_d|^2 + m_S^2|S|^2
+ [\lambda A_{\lambda}SH_uH_d+\frac{1}{3}\kappa A_{\kappa}S^3+\textrm{h.c.} ], 
\label{eq:HpotS}
\ea
where $\bar g = \sqrt{g^2+g^{\prime 2}}$ with $g$ and $g^{\prime}$
being the gauge couplings of $SU(2)_L$ and $U(1)$ interactions
respectively, and adopting the notation $H_uH_d \equiv
\epsilon_{\alpha \beta}(H_u)^{\alpha}(H_d)^{\beta}=H_u^+H_d^--H_u^0H_d^0$.  
The first two terms, $V_F$ and $V_D$, are the $F$ and $D$ terms
derived from the superpotential in the usual way.  We also include
additional soft supersymmetry breaking terms, $V_{\rm soft}$, by adding
all the forms in the superpotential with arbitrary (dimensionful)
coefficients\footnote{These definitions differ from those in
Ref.\cite{ellis} w.r.t. the sign of $A_{\kappa}$}, and soft mass
terms for the fields. This leads to the inclusion of the extra
parameters: $A_{\lambda}$, $A_{\kappa}$, $m_{H_u}$, $m_{H_d}$ and
$m_S$.

The parameter combinations $\lambda A_{\lambda}$ and $\kappa
A_{\kappa}$ may always be taken to be real and positive since their
complex phases may be absorbed into global redefinitions of $H_uH_d$
and $S$. However, $\lambda$ and $\kappa$ (and therefore $A_{\lambda}$
and $A_{\kappa}$) may be complex. Significantly, if the combination
$\lambda\kappa^*$ is complex then CP will be violated at tree--level
in the Higgs sector. This is in contrast to the MSSM where tree--level
CP conservation is guaranteed by the structure of the Higgs
potential. In this paper we will adopt the usual convention of
assuming that all four parameters are real. For studies of the
CP--violating NMSSM see Ref.\cite{CP_violating _nmssm}.

While the essential new elements of the Higgs sector can be elaborated
at the tree-level, one-loop contributions to the Higgs masses and
couplings from top and stop loops are included in the quantitative
analyses by introducing the term $V_{\rm loop}$ into the potential,
with~\cite{vloop}
\be
\langle V_{\rm loop}\rangle = \frac{3}{32 \pi^2} \left[ 
 m_{\tilde{t}_1}^4 \left( \log \frac{m_{\tilde{t}_1}^2}{Q^2}-\frac{3}{2} \right)
+m_{\tilde{t}_2}^4 \left( \log \frac{m_{\tilde{t}_2}^2}{Q^2}-\frac{3}{2} \right)
-2m_t^4 \left( \log \frac{m_t^2}{Q^2}-\frac{3}{2} \right) \right].
\label{eq:Vloop}
\ee 
The dependence on the Higgs fields is implicitly contained in the top,
$m_t$, and stop masses, $m_{\tilde{t}_{1,2}}$, which are dependent, in
total or partially, on the Higgs VEVs. For the stop masses, we use the
tree--level relation,
\be
m^2_{\tilde{t}_{2/1}}=\frac{1}{2} \left( m_Q^2 + m_U^2 + 2 m_t^2 \pm
\sqrt{(m_Q^2-m_U^2)^2+4m_t^2X_t^2} \right), 
\ee 
where we set $m_Q=m_U=M_{SUSY}=1$ TeV, the running top quark mass
$m_t=165$~GeV and, as an example, assume maximal mixing, $X_t=\sqrt{6}
M_{SUSY}$. The two physical stop masses are given by
$m_{\tilde{t}_1}=789$~GeV and $m_{\tilde{t}_2}=1196$~GeV for these
parameters. The renormalization scale, $Q$, is taken to be the running
top quark mass.  For the present analysis one-loop expressions for the
masses and couplings suffice. The explicit expressions for the
one-loop corrections to the mass matrices were adopted from
Ref.\cite{roman}.\\

The ``MSSM limit'' can be approached smoothly by letting $\lambda$ and
$\kappa \to 0$ on a linear trajectory (i.e.~$\kappa/\lambda$
constant), while keeping the $\mu$ parameter $\mu=\lambda
v_s/\sqrt{2}$ and the parameters $A_{\lambda}$ (which plays the role
of $B$ in the MSSM) and $A_{\kappa}$ fixed. In this limit the Higgs
singlet field decouples from the system completely, and we regain the
doublet Higgs sector of the MSSM in its canonical structure.\\

The structure of the vacuum can be simplified by making a 
$SU(2)_L \times U(1)_Y$ gauge transformation to choose 
$\langle H_u^- \rangle = 0$. In order to ensure that 
$\langle H_d^- \rangle =0$ is a suitable choice of vacuum, 
i.e.~that the vacuum is uncharged, one also requires
that it be a stable minimum\footnote{The requirement $\langle
\frac{\partial V}{\partial H_d^-} \rangle=0$ is trivially satisfied for
$\langle H_d^- \rangle =0$.} with respect to small perturbations in $H_d^-$, 
corresponding to a positive charged Higgs mass-squared:
\be \langle \frac{\partial^2 V}{\partial H_d^- \, 
\partial H_d^{- \, *}} \rangle = m_{H_d}^2+\frac{1}{2}\lambda^2 |v_s|^2
+\frac{1}{8}g^2(|v_u|^2+|v_d|^2)-\frac{1}{8}g^{\prime
2}(|v_u|^2-|v_d|^2)>0, \label{eq:stab} \ee 
where $v_u$ and $v_d$ are the VEVs of $H_u^0$ and $H_d^0$
respectively, multiplied by a factor of $\sqrt{2}$. The resulting
vacuum has been studied in Ref.\cite{ellis}, where it was shown
that there are three degenerate vacua which differ in the complex
phases of $\langle S
\rangle$ and $\langle H_d^0 \rangle$. Here, we will restrict ourselves
to a discussion of the vacuum where all the remaining VEVs are real, as
described by,
\be 
        \langle H_d \rangle = \frac{1}{\sqrt{2}} {v_d \choose 0}, 
\qquad  \langle H_u \rangle = \frac{1}{\sqrt{2}} {0 \choose v_u}, 
\qquad  \langle S \rangle = \frac{1}{\sqrt{2}} v_s, \label{eq:potmin} \ee
with $v_s$, $v_u$, and $v_d$ real and positive.

For this vacuum to be a local minimum, we obtain three relations,
linking the three soft mass parameters to the three VEVs of the Higgs
fields:
\ba
m_{H_d}^2&=& \frac{1}{8} {\bar g}^2 (v_u^2-v_d^2) -\frac{1}{2} \lambda^2 v_u^2
 + \frac{1}{2}(\sqrt{2} A_{\lambda}  +\kappa v_s) \lambda v_s \frac{v_u}{v_d} 
- \frac{1}{2} \lambda^2 v_s^2, \label{eq:mind}\\
m_{H_u}^2&=& \frac{1}{8} {\bar g}^2 (v_d^2-v_u^2) -\frac{1}{2} \lambda^2 v_d^2 
+ \frac{1}{2}(\sqrt{2} A_{\lambda}  +\kappa v_s) \lambda v_s \frac{v_d}{v_u} 
- \frac{1}{2} \lambda^2 v_s^2, \label{eq:minu} \\
m_S^2 &=& -\kappa^2v_s^2 -\frac{1}{2}\lambda^2v^2 +\kappa\lambda v_uv_d
+\frac{1}{\sqrt{2}} \lambda A_{\lambda} \frac{v_uv_d}{v_s}
-\frac{1}{\sqrt{2}} \kappa A_{\kappa}v_s; \label{eq:mins}
\ea
as usual, we have written $v\equiv \sqrt{v_u^2+v_d^2}$.  The {\it
local} stability of this vacuum is ensured by allowing only positive
squared masses of the physical fields (i.e. the mass eigenstates),
which leads to useful constraints on the parameters of the
potential.\\

Proving this to be a {\it global} minimum of the vacuum is beyond the
scope of this study. However, we consider all possible neutral CP--even
vacuum states and ensure that the physical vacuum is of lower energy: \\ 

({\it i}) First of all, it is clear that $\langle H_u^0 \rangle =
\langle H_d^0 \rangle = \langle S \rangle = 0$ is also a 
stationary point. It must be ensured that this symmetric minimum is
not preferred to the physical vacuum with non-zero fields. This is
possible in two ways. For the symmetric vacuum to be a stable minimum
requires all of $m_{H_u}^2$, $m_{H_d}^2$ and $m_S^2$ to be
positive\footnote{This constraint is different from the corresponding
condition in the MSSM since $\langle S \rangle =0$, removing the
effective $\mu$--term.}. Therefore, choosing the parameters such that
one of the soft masses is negative, is sufficient. However, while {\it
sufficient}, this constraint is not {\it necessary}. Alternatively,
one may check that $\langle V \rangle <0$ for the physical vacuum,
making it a lower energy state than the symmetric vacuum for which
$\langle V \rangle =0$. This constraint places an upper bound on the
parameter $A_{\lambda}$. \\

({\it ii}) The only other neutral CP-even vacua which could be
problematic are those where two of $\langle H_d^0 \rangle$, $\langle
H_u^0 \rangle$ and $\langle S \rangle$ are zero and the other
non-zero. The requirement that the resulting vacuum energy $\langle V
\rangle$ is greater for these points than for the physical vacuum can
be fulfilled in the major part of the appropriate parameter space.
Vacua where only one of $\langle H_d^0 \rangle$, $\langle H_u^0
\rangle$, $\langle S \rangle$ is zero lead to an over-constrained
system, so that these vacua can only be realized for very specific
choices of the parameters, and may be safely ignored. \\

The NMSSM Higgs potential is automatically bounded from below for
non-zero $\kappa$. The two terms in $V_F$ contain contributions which
are quartic in the usual neutral Higgs fields, $H_u$, $H_d$, and in
the new scalar, $S$, and will ensure that the potential is bounded
from below.

\subsection{The Mass Matrices}

From the potential, the Higgs mass matrices and subsequently the mass
eigenstates can be derived. After shifting the Higgs fields to the
minimum of the potential (given by Eqn.(\ref{eq:potmin})), they are
rotated by an angle $\beta$ in order to isolate the zero mass
Goldstone states, $G$, which are absorbed by the $W$ and $Z$ bosons to
provide their masses.

For the charged fields these redefinitions can be written:
\be \begin{array}{rcl}
H_d^-&=&H^- \sin \beta - G^- \cos \beta, \\
H_u^+&=&H^+ \cos \beta + G^+ \sin \beta, 
\end{array} \label{eq:chdef} \ee
where $G^-=G^{+\,*}$ and $H^-=H^{+\,*}$. For the imaginary and real
field components we have
\be \begin{array}{rcl}
\Im m\, H_d^0&=& (P_1 \sinb - G^0 \cosb)/\sqrt{2}, \\
\Im m\,H_u^0&=& (P_1 \cosb + G^0 \sinb)/\sqrt{2}, \\
\Im m\,S&=& P_2/\sqrt{2}, \end{array} \label{eq:pdef} \ee
and
\be \begin{array}{rcl}
\Re e\, H_d^0&=& (- S_1 \sinb + S_2 \cosb + v_d)/\sqrt{2}, \\
\Re e\, H_u^0&=& (\phantom{-} S_1 \cosb + S_2 \sinb + v_u)/\sqrt{2}, \\
\Re e\, S&=& (S_3+v_s)/\sqrt{2}, \\ 
\end{array} \label{eq:sdef}
\ee
respectively. Choosing $\tan \beta = v_u/v_d$ the zero-mass Goldstone
modes decouple, and the resulting potential has terms for the non-zero 
mass modes given by
\be V_{\rm mass} = 
M_{H^{\pm}}^2  H^+ H^- 
+ \frac{1}{2} \,(P_1 \,\,\, P_2)\, M_-^2 
\left( \begin{array}{ll} P_1 \\ P_2 \end{array} \right)
+ \frac{1}{2} \,(S_1 \,\,\, S_2 \,\,\, S_3)\, M_+^2 
\left( \begin{array}{lll} S_1 \\ S_2 \\ S_3 \end{array} \right).
\label{eq:massterms} \ee

The \underline{charged} fields $H^{\pm}$ are already physical mass
eigenstates with tree-level masses given by \vspace*{0.6cm} \\ $(I):$
\vspace*{-0.6cm}
\be M_{H^{\pm}}^2 = M_A^2 +M_W^2- \frac{1}{2} (\lambda v)^2, 
\label{eq:chmass} \ee
where
\be M_A^2= \frac{\lambda v_s}{\sin 2 \beta} 
\left( \sqrt{2} A_{\lambda} + \kappa v_s \right). \label{eq:ma} \ee

In contrast, the \underline{pseudoscalar} and \underline{scalar} fields, 
$P_i$ ($i=1,\,2$) and
$S_i$ ($i=1,\,2,\,3$) respectively, are not yet mass
eigenstates. Their mass matrices, $M_-^2$ and $M_+^2$, must be further
rotated to a diagonal basis corresponding to the physical mass
eigenstates.  Using the minimization conditions,
Eqns.(\ref{eq:mind}--\ref{eq:mins}), the tree--level CP--odd matrix
has entries, \vspace*{0.6cm} \\ $(II):$
\vspace*{-0.95cm}
\ba 
M_{-\,11}^2&=&M_A^2, \label{eq:m11odd} \\
M_{-\,12}^2&=&\frac{1}{2} (M_A^2\sin2\beta-3\lambda\kappa v_s^2) \cot \beta_s,\\
M_{-\,22}^2&=& \frac{1}{4}(M_A^2\sin2\beta+3\lambda\kappa v_s^2)\cot^2 \beta_s \sin2\beta
-3\kappa v_sA_{\kappa}/\sqrt{2}.
\label{eq:m22odd} \ea
while the tree--level CP--even matrix is given by, \vspace*{0.6cm} \\ $(III):$
\vspace*{-0.95cm}
\ba
M_{+\,11}^2&=& M_A^2 + (M_Z^2-\half (\lambda v)^2) \sin^2 2 \beta,
\label{eq:m11even}\\
M_{+\,12}^2&=&- \frac{1}{2} (M_Z^2-\half (\lambda v)^2) \sin 4 \beta, \\
M_{+\,13}^2&=&
-\frac{1}{2}(M_A^2\sin2\beta+\lambda \kappa v_s^2)\cot \beta_s \cos 2\beta,\\
M_{+\,22}^2&=&M_Z^2\cos^2 2 \beta + \half (\lambda v)^2 \sin^2 2 \beta,
\label{eq:m22even}\\ 
M_{+\,23}^2&=&\frac{1}{2}(2\lambda^2v_s^2-M_A^2\sin^22\beta-\lambda\kappa 
v_s^2\sin2\beta)\cot \beta_s,\\
M_{+\,33}^2&=&\frac{1}{4}M_A^2\sin^22\beta \cot^2 \beta_s+2\kappa^2v_s^2
+\kappa v_s A_{\kappa}/\sqrt{2}-\frac{1}{4}\lambda\kappa v^2\sin2\beta,
\label{eq:m33even} \ea
Besides the the usual notation for $\tan \beta$, 
we have introduced the auxiliary ratio $\tan \beta_s
\equiv v_s/v$.  The mass parameter $M_A^2$ is {\it defined} to be the
top--left entry of the CP--odd squared mass matrix,
c.f.~Eqn.(\ref{eq:m11odd}), which is positive in the physical
parameter ranges analyzed later. It becomes the mass of the heavy
pseudoscalar Higgs boson in the ``MSSM limit''. Parallel to the MSSM,
$M_A$ replaces the soft parameter $A_{\lambda}$, and at tree--level is
given by Eqn.(\ref{eq:ma}). \\

\noindent
\underline{\it Higgs Mass Spectrum} \\[-0,5cm]

  The \underline{charged Higgs mass} has been noted in
Eqn.(\ref{eq:chmass}). The condition for stability of the vacuum in
the $H_d^-$ direction, Eqn.(\ref{eq:stab}), is simply equivalent to
the positivity of $M^2_{H^{\pm}}$.

The above pseudoscalar and scalar mass matrices,
Eqns.~(\ref{eq:m11odd}--\ref{eq:m33even}), do not lend themselves
easily to obtaining analytic expressions for the physical Higgs
masses. However, reasonably simple expressions can be found by
performing a systematic expansion for large $\tan \beta$ and large
$M_A$ [on the generic electroweak scale].  This is outlined in detail
in the appendix. This approximation works extremely well, even for
moderate values of $\tan \beta$, and it may therefore be used to shed
light on the behaviour of the Higgs masses as the other parameters are
varied.

This approximation may be used to simplify the expressions for the
tree-level \underline{CP-odd masses}, giving,
\ba
M_{A_2}^2 &\approx& M_A^2 \, (1+\frac{1}{4} \cot^2 \beta_s \sin^2 2\beta),\\
M_{A_1}^2 &\approx& -\frac{3}{\sqrt{2}} \kappa v_s A_{\kappa}, \label{eq:ma1app}
\ea
while the tree-level physical masses of the \underline{CP--even Higgs
bosons} are, in this approximation,
\ba
M_{H_3}^2 &\approx& M_A^2 (1+\frac{1}{4} \cot^2 \beta_s \sin^2 2\beta), \\
M_{H_{2/1}}^2 &\approx& \frac{1}{2} \left\{ M_Z^2
+\frac{1}{2} \kappa v_s (4\kappa v_s+\sqrt{2}A_{\kappa}) \right. \nonumber \\
&&\quad \quad \left. \pm \sqrt{ \left[M_Z^2
-\frac{1}{2} \kappa v_s (4\kappa v_s+\sqrt{2}A_{\kappa}) \right]^2
+\cot^2 \beta_s \left[2\lambda^2v_s^2-M_A^2\sin^22\beta \right]^2}
\right\}. \label{eq:approx12}
\ea
The physical mass eigenstates $A_i$ and $H_i$ are labelled in
ascending order of mass. 

The heavy CP-odd Higgs boson, $A_2$, is approximately degenerate with
the heaviest CP-even Higgs boson, $H_3$, and the charged Higgs bosons,
cf. Eqn.(\ref{eq:chmass}). One of the lighter CP-even tree-level
masses will be of the order of $M_Z$ while the scale of the other is
set by $\sim \kappa v_s$ for $\kappa$ and $\lambda$ sufficiently below
unity. Finally the lightest pseudoscalar mass grows as $\kappa$, $v_s$
and $A_{\kappa}$ are increased, with a negative value for $A_{\kappa}$
being preferred.\\

\noindent
\underline{\it Range of Parameters} \\[-0.5cm]

This above solution also allows limits to be placed on the Higgs
potential parameters. First of all, the lightest two scalar Higgs
boson masses respect a sum rule:
\be 
M_{H_1}^2+M_{H_2}^2 \approx M_Z^2
+\frac{1}{2} \kappa v_s (4\kappa v_s+\sqrt{2}A_{\kappa}).
\ee
The right-hand side, and thus the sum of the two lightest scalars, is
independent of the coupling $\lambda$ and $M_A^2$.  As a result of
this sum rule, the second lightest scalar Higgs boson mass is
maximized as the lightest approaches zero.

Furthermore, the smallest eigenvalue of a matrix is smaller than its
smallest diagonal entry and similarly the largest eigenvalue is larger
than its largest diagonal entry. Together with the above sum rule,
this leads to the mass constraints at tree-level (i.e. modified by
radiative corrections in parallel to the MSSM):
\ba  &M_{H_1}^2 
&\lesssim {\rm min}\{M_Z^2, 
\frac{1}{2} \kappa v_s (4 \kappa v_s + \sqrt{2} A_{\kappa})\}, \\
{\rm max}\{M_Z^2, 
\frac{1}{2} \kappa v_s (4 \kappa v_s + \sqrt{2} A_{\kappa})\}
\lesssim 
&M_{H_2}^2 &\lesssim M_Z^2 +
\frac{1}{2} \kappa v_s (4 \kappa v_s + \sqrt{2} A_{\kappa}).
\ea

A further constraint is found by exploiting the condition that the
mass-squared of the lightest Higgs boson, given in
Eqn.(\ref{eq:approx12}), is greater than zero. This in turn gives a
restriction on the allowed values of $M_A$. At tree--level, again to
the accuracy of the approximate solution, this is,
\be
M_A^2 \, \lessgtr \,
\frac{2 \lambda^2 v_s^2}{\sin^2 2 \beta}
\pm 
\frac{M_Z \tan \beta_s}{\sin^22\beta} 
\sqrt{\frac{\kappa v_s}{2} (4 \kappa v_s +\sqrt{2}A_{\kappa})},
\label{eq:malim}
\ee
which generalises the corresponding constraint for zero $\kappa$ in
Ref.~\cite{kill_tadpoles2}.  In the theoretically preferred scenarios
where $\kappa$ is small, the second term above is small compared to
the first term, and we find that the value of $M_A$ is constrained to
lie in a narrow bracket, not too far from the value $\sqrt{2} \lambda
v_s / \sin 2 \beta = 2\mu/\sin 2\beta$ which is approximately $M_A
\approx \mu \tan \beta$ for medium to large values of $\tan \beta$. As
can be seen from Eqn.(\ref{eq:approx12}), this is also the approximate
value of $M_A$ for which $M_{H_1}$ is maximal (and $M_{H_2}$
minimal). The constraint on $M_A$ becomes stronger as $\kappa$ becomes
smaller but $v_s$ is kept fixed. The mass parameter $M_A$ is
\underline{not constrained} in the proper ``MSSM limit'' defined
earlier with $\lambda v_s$ and $\kappa v_s$ fixed for $\lambda$ and
$\kappa \to 0$.

Finally, one may also gain some insight into the allowed range of the
soft SUSY breaking parameter $A_{\kappa}$. Requiring the lightest
scalar and pseudoscalar mass-squareds be positive leads to upper and
lower constraints on $A_{\kappa}$. At tree--level, these are given by,
\be
-2\sqrt{2} \kappa v_s \lesssim A_{\kappa} \lesssim 0. \label{eq:aklim}
\ee
The lower bound is derived from the requirement that the {\it maximum}
value of the lightest scalar mass-squared be positive, which is
realized for the central value of $M_A$ given in
Eqn.(\ref{eq:malim}). For values of $M_A$ deviating from this, the
lower bound will become stronger.

One should bear in mind that most of these constraints are based upon
the approximate solution as described in the appendix, and will become
unreliable in regions of parameter space where the approximate
solution breaks down. Nevertheless they provide the proper analytical
understanding of the exact numerical results presented in Sec.(3).

\subsection{Couplings with the $Z$ boson}

The couplings of the Higgs bosons to the $Z$ boson are given by the
Lagrangian,
\ba
{\cal L}_{ZAH} &=& \frac{\bar g}{2} M_Z Z_{\mu} Z^{\mu} S_2 
+ \frac{\bar g}{2} Z_{\mu} \left[ S_1(\partial^{\mu} P_1)
-P_1(\partial^{\mu}S_1) \right] \nn \\ 
&& \qquad +\left( eA_{\mu} + \frac{\bar g}{2} Z_{\mu} \cos 2 \theta_W \right)
i \left[ H^- (\partial^{\mu} H^+) - H^+ (\partial^{\mu} H^-) \right],
\ea
where $Z^{\mu}$ and $A^{\mu}$ are the $Z$ boson and photon fields
respectively, $\theta_W$ is the weak mixing angle and the Higgs fields
have been defined in Eqns.(\ref{eq:chdef}--\ref{eq:sdef}). The simple
form of this Lagrangian arises from the rotation made on the CP--even
Higgs states. This leads to $S_2$ being the only field with a coupling
of the form $ZZS_i$, while the scalar and pseudoscalar fields are
coupled jointly to $Z$ only in the form $ZS_1P_1$. The couplings of
the extra scalar and pseudoscalar Higgs bosons to the $Z$ are due only
to their mixing with the Higgs doublet degrees of freedom.  We define
this mixing of the CP--odd Higgs bosons via a rotation by an angle
$\theta_A$ which transforms the fields to their physical mass
eigenstates:
\be
\left( \begin{array}{c} A_2 \\ A_1 \end{array} \right)
=  \left( \begin{array}{cc} \cos \theta_A & \sin \theta_A \\ 
-\sin \theta_A & \cos \theta_A \end{array} \right)
\left( \begin{array}{c} P_1 \\ P_2 \end{array} \right). 
\ee
At tree--level, the mixing angle, $\theta_A$, is given by, 
\be
\tan \theta_A = \frac{M_{-\,12}^2}{M_{-\,11}^2-M_{A_1}^2}=
\frac{1}{2} \cot \beta_s
\frac{M_A^2 \sin 2\beta -3 \lambda\kappa v_s^2}{M_A^2-M_{A_1}^2}.
\label{eq:tantha} \ee

Similarly, the more complicated scalar mixing is defined via an
orthogonal $3 \times 3$ rotation matrix $O$ such that,
\be 
\left( \begin{array}{c} H_3 \\ H_2 \\ H_1 \end{array} \right) = 
O \left( \begin{array}{c} S_1 \\ S_2 \\ S_3 \end{array} \right).
\label{eq:HtoS} \ee
Our notation is such that the fields $A_i$, $i=1,2$, and $H_i$,
$i=1,2,3$ are the mass eigenstates of the pseudoscalar and scalar
Higgs sectors respectively, with the numerical suffix denoting their
mass hierarchy in ascending order (e.g.~$A_1$ ($H_1$) is the lightest
pseudoscalar (scalar) Higgs boson). [The ordering of the states $S_i$
has been introduced such that $S_3$ is the decoupling singlet.]

It is more useful to discuss the couplings of these Higgs bosons to
the $Z$ in terms of normalized couplings. To this end, we define the
normalized CP--even couplings by dividing out the associated SM
coupling and for the $ZA_iH_j$ couplings we similarly divide by $\bar
g/2=\sqrt{g^2+g^{\prime \, 2}}/2$, so that,
\be
\mathcal{G}_{ZZH_i} \equiv \frac{g^{\rm NMSSM}_{ZZH_i}}{g^{\rm SM}_{ZZH}},
\qquad \mathcal{G}_{ZA_iH_j} \equiv
\frac{g^{\rm NMSSM}_{ZA_iH_j}}{\bar g/2}.
\ee
The couplings may then be written directly in terms of the mixing
matrices, in an obvious notation, according to,
\be \mathcal{G}_{ZZH_i} = O_{H_i S_2}, \label{eq:rzzh} \ee
and 
\be \mathcal{G}_{ZA_1H_i}=-O_{H_i S_1} \sin \theta_A,
\qquad \mathcal{G}_{ZA_2H_i}=O_{H_i S_1} \cos \theta_A. \ee
For these normalized couplings the orthogonality of the mixing matrices
lead to the constraints,
\be
\sum_i \mathcal{G}^2_{ZZH_i} = 1, \quad \sum_{ij} \mathcal{G}^2_{ZA_iH_j} = 1, 
\quad \sum_i \mathcal{G}_{ZZH_i}\mathcal{G}_{ZA_jH_i} =0.
\ee
Finally, the couplings of a pseudoscalar and a scalar to a $Z$ boson
are not all independent. The ratios of the two pseudoscalar couplings,
$A_1$ and $A_2$, are independent of the scalar $H_i$:
\be 
\frac{\mathcal{G}_{ZA_1H_1}}{\mathcal{G}_{ZA_2H_1}} = 
\frac{\mathcal{G}_{ZA_1H_2}}{\mathcal{G}_{ZA_2H_2}} = 
\frac{\mathcal{G}_{ZA_1H_3}}{\mathcal{G}_{ZA_2H_3}} = -\tan \theta_A.
\label{eq:couprel}
\ee \\

In the approximate solution, where $1/M_A$ and $1/\tan \beta$ are
regarded as small quantities, the orthogonal matrix $O$ is given by
Eqn.(\ref{eq:Uapprox}) in the appendix. Inserting this into
Eqn.(\ref{eq:rzzh}), gives approximate expressions for the
scalar-$Z$-$Z$ couplings,
\be
\mathcal{G}_{ZZH_1} \approx -\sin \theta_H, \qquad\qquad \mathcal{G}_{ZZH_2} \approx \cos \theta_H,
\ee \be
\mathcal{G}_{ZZH_3} \approx M_{12}^2/M_{11}^2 \approx \frac{1}{2 M_A^2} \left[ \half (\lambda v)^2 -M_Z^2
\right]\sin 4\beta,
\ee
where $\tan \theta_H$ is given by (see Eqn.(\ref{eq:tanthh})),
\be
\tan \theta_H \approx 
\frac{1}{2} \cot \beta_s
\frac{2 \lambda^2 v_s^2 - M_A^2 \sin^2 2\beta
- \kappa \lambda v_s^2 \sin 2 \beta}
{M_Z^2\cos^2 2 \beta + \half \lambda^2 v^2 \sin^2 2 \beta - M_{H_{1}}^2}.
\ee

Similarly, for the $Z$-pseudoscalar-scalar couplings we have,
\ba
\mathcal{G}_{ZA_1H_1}&\approx&
\frac{1}{4 M_A^2}  \left[M_A^2 \sin2\beta +
\kappa \lambda v_s^2 \right] \cot \beta_s \cos 2 \beta \sin \theta_A, \\
\mathcal{G}_{ZA_1H_2}&\approx&-\frac{1}{2M_A^2} \left[ \half (\lambda v)^2 -M_Z^2 \right]\sin 4\beta \sin \theta_A, \\
\mathcal{G}_{ZA_1H_3}&\approx&-\left[1+
\frac{1}{4 M_A^2} \left[M_A^2 \sin2\beta +
\kappa \lambda v_s^2 \right] \cot \beta_s \cos 2 \beta \right] \sin \theta_A,
\ea
and, following from the relations Eqn.(\ref{eq:couprel}),
\ba
\mathcal{G}_{ZA_2H_1}&\approx&
-\frac{1}{4 M_A^2} \left[M_A^2 \sin2\beta +
\kappa \lambda v_s^2 \right] \cot \beta_s \cos 2 \beta \cos \theta_A, \\
\mathcal{G}_{ZA_2H_2}&\approx&\frac{1}{2M_A^2} \left[ \half (\lambda v)^2 -M_Z^2 \right]\sin 4\beta \cos \theta_A, \\
\mathcal{G}_{ZA_2H_3}&\approx&\left[1+
\frac{1}{4 M_A^2} \left[M_A^2 \sin2\beta +
\kappa \lambda v_s^2 \right] \cot \beta_s \cos 2 \beta \right] \cos \theta_A,
\ea
where $\tan \theta_A$ is given by Eqn.(\ref{eq:tantha}).

\section{NMSSM Higgs boson scenarios}

\subsection{The parameters} 

At tree--level the NMSSM Higgs sector described above has six free
parameters (apart from the overall scale, set later by the $Z$ boson
mass). The Higgs potential, Eqns.(\ref{eq:Hpot}--\ref{eq:HpotS}),
contained seven parameters: $\lambda$ and $\kappa$ from the
superpotential and $A_{\lambda}$, $A_{\kappa}$, $m_{H_d}$, $m_{H_u}$
and $m_S$ from the soft supersymmetry breaking terms. The field values
at the minimum of the potential, $v_u$, $v_d$ and $v_s$, are fixed by
these parameters according to Eqns.(\ref{eq:mind}--\ref{eq:mins}). The
structure of electroweak symmetry breaking allows us to remove one
(combination) of these VEVs in favour of the known electroweak scale
$v \equiv \sqrt{v_u^2+v_d^2}=246$ GeV (or equivalently, the $Z$ boson
mass, $M_Z$), defining the overall mass scale. Finally, after
introducing $\tan \beta \equiv v_u/v_d$ and writing $A_{\lambda}$ in
terms of the heavy Higgs mass scale $M_A$, we are therefore left with
the six free parameters, $\lambda$, $\kappa$, $v_s$, $\tan \beta$,
$M_A$, and $A_{\kappa}$ ($v_s$ may be expressed in some places in
terms of the effective $\mu$ parameter,
\mbox{$v_s=\sqrt{2} \mu/\lambda$}).  The spectrum of the NMSSM is expected to
show a strong dependence on these six free parameters, which we shall
now analyze in detail.

Including higher orders introduces extra parameters such as the top
and stop (s)quark masses.  We choose to fix these extra parameters at
reasonable values and not vary them. \\

\underline{$\lambda$ and $\kappa$} \\[-0.4cm]

Requiring a weak coupling of the fundamental fields,
i.e.~field-theoretic perturbativity, in the entire range between the
electroweak and GUT scales restricts the range of values for the
couplings $\lambda$ and $\kappa$ at the electroweak scale.  The
renormalization group equations for $\lambda$, $\kappa$ and the top
Yukawa coupling $h_t$ form a closed set together with those of the
gauge couplings. They are given by~\cite{ds84,falck,NT01}
\ba
16\pi^2\,\frac{d g_i^2}{dt} &=& b_ig_i^2, \label{eq:rgeg} \\
16\pi^2\,\frac{d h_t^2}{dt} &=& h_t^2 \left[ \lambda^2+6h_t^2
-\frac{16}{3}g_3^2 - 3 g_2^2 -\frac{13}{15} g_1^2 \right], \label{rgeh} \\
16\pi^2\,\frac{d \lambda^2}{dt} &=& \lambda^2 
\left[ 4\lambda^2+2\kappa^2+3h_t^2
- 3g_2^2 -\frac{3}{5}g_1^2 \right], \label{eq:rgel} \\
16\pi^2\,\frac{d \kappa^2}{dt} &=& 6\kappa^2 
\left[\lambda^2+\kappa^2 \right], \label{eq:rgek}
\ea
where $b_1=33/5$, $b_2=1$, $g_1=\sqrt{5/3}\,g'$, $g_2=g$ and $t= \log
(Q^2/M_{\rm GUT}^2)$.

\begin{figure}[t]
\mbox{\hspace*{-0.5cm}\includegraphics[clip=true,width=9.5cm]{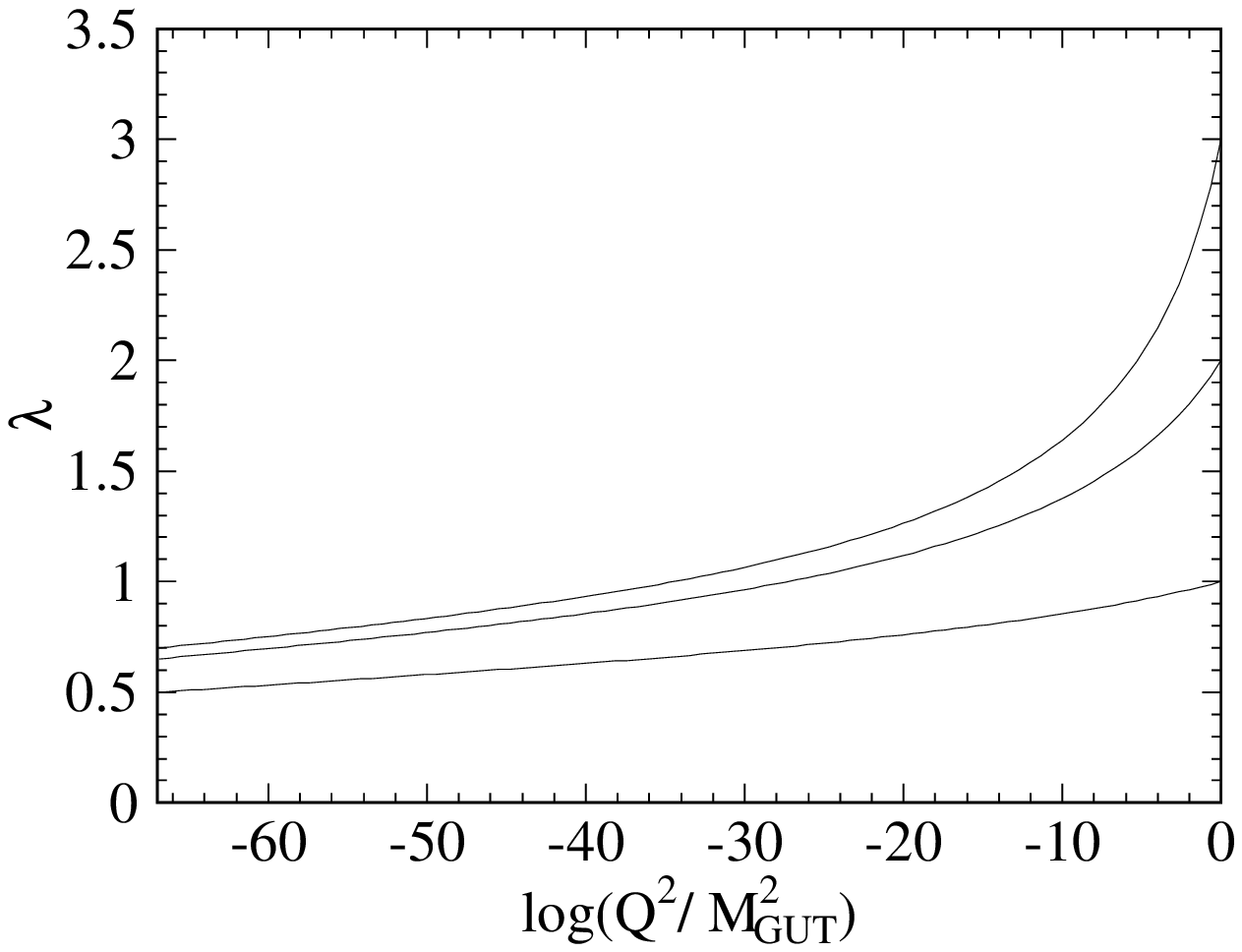}
\hspace*{-1cm}\includegraphics[clip=true,width=9.5cm]{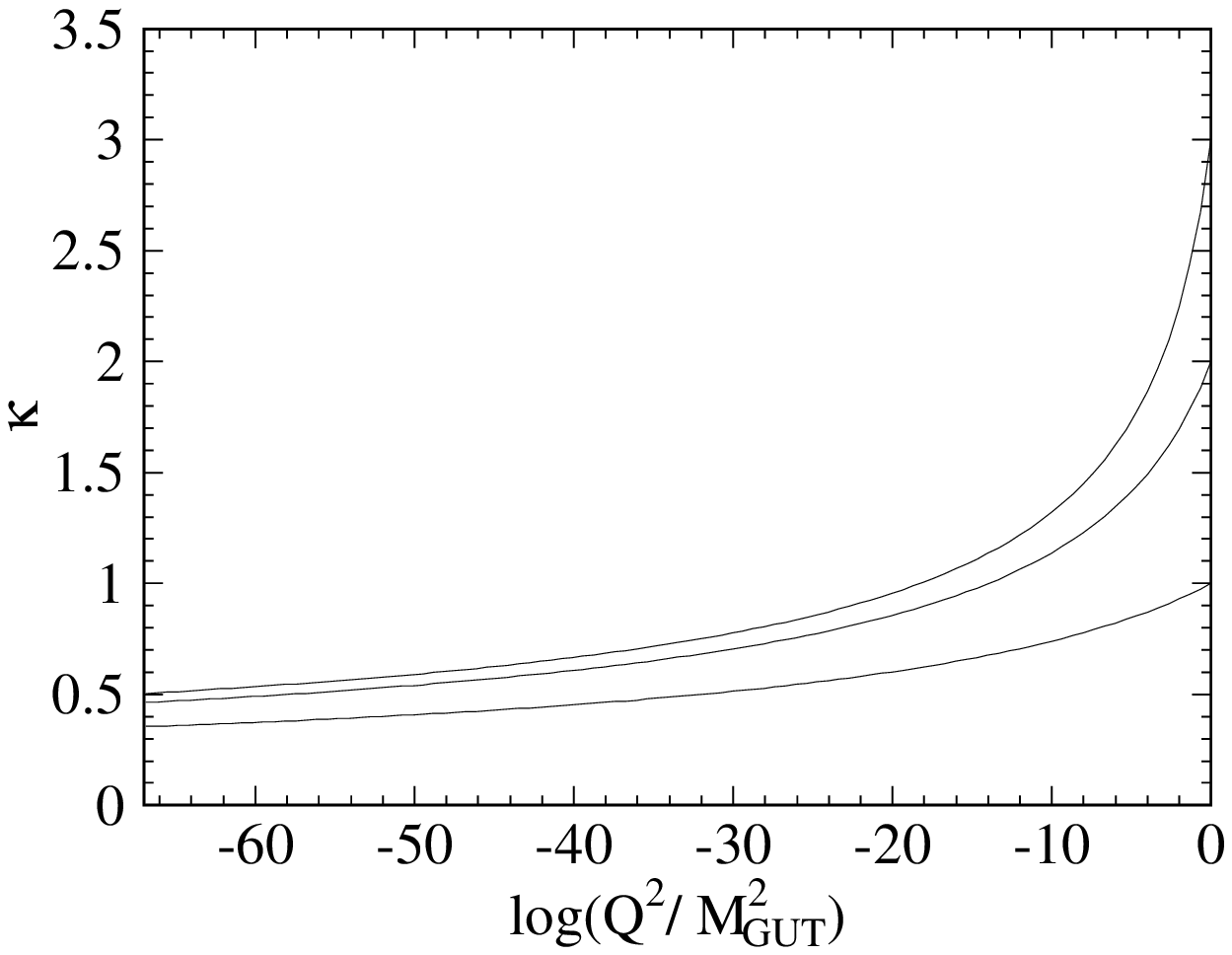}}
\caption{\it The dependence of $\lambda$ (left) and 
         $\kappa$ (right) on renormalization scale, $Q$, for the top
         Yukawa coupling \mbox{$h_t(M_{\rm GUT})=0.8$}. The different
         curves represent different values of $\lambda$ and $\kappa$
         at the GUT scale; {\it left:} $\lambda=1$, $2$ and $3$, with
         $\kappa=1$; {\it right:} $\kappa=1$, $2$ and $3$, with
         $\lambda=1$.}
\label{fig:lkrun} 
\end{figure}
\begin{figure}[h]
\mbox{\hspace*{-0.5cm}\includegraphics[clip=true,width=9.5cm]{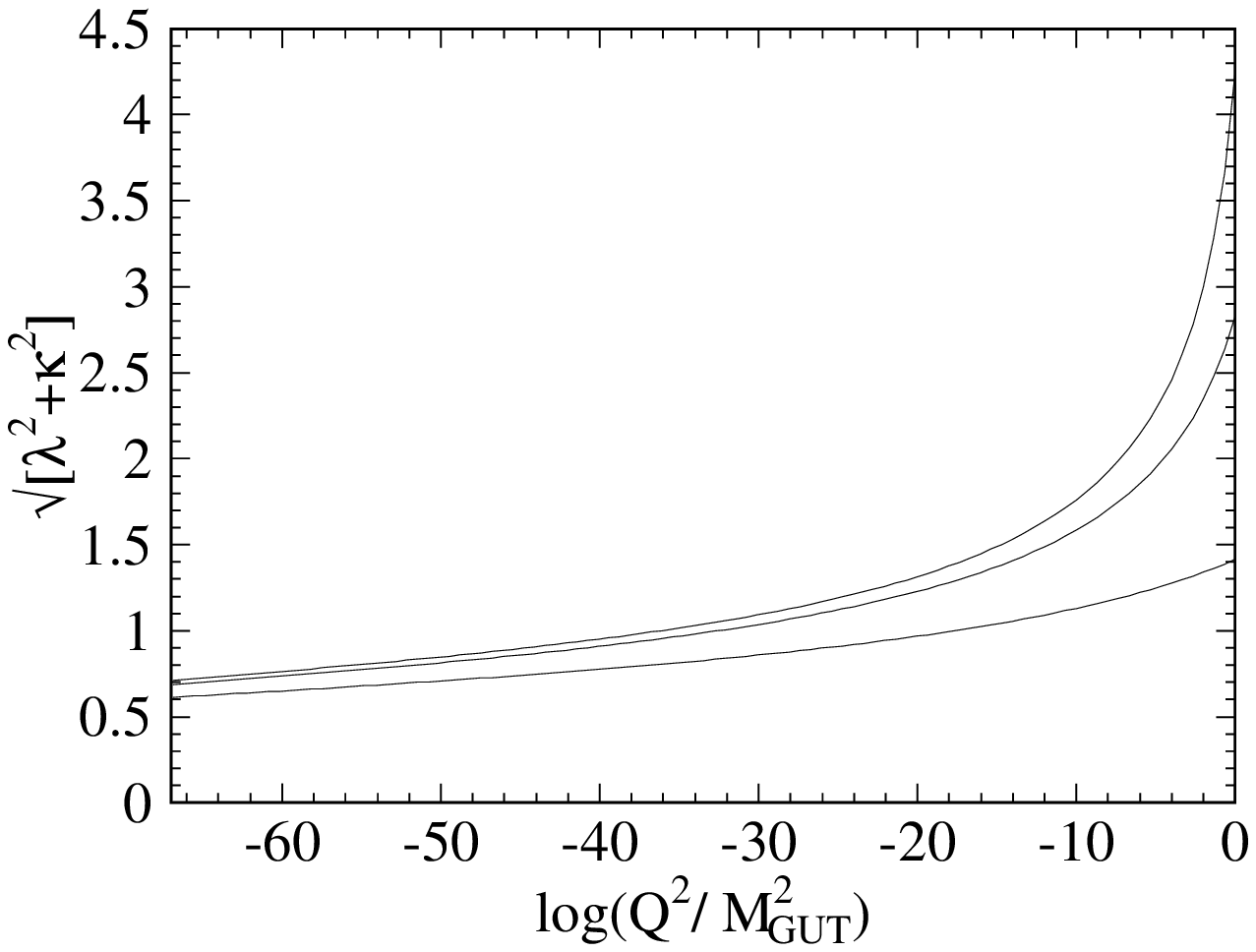}
\hspace*{-1cm}\includegraphics[clip=true,width=9.5cm]{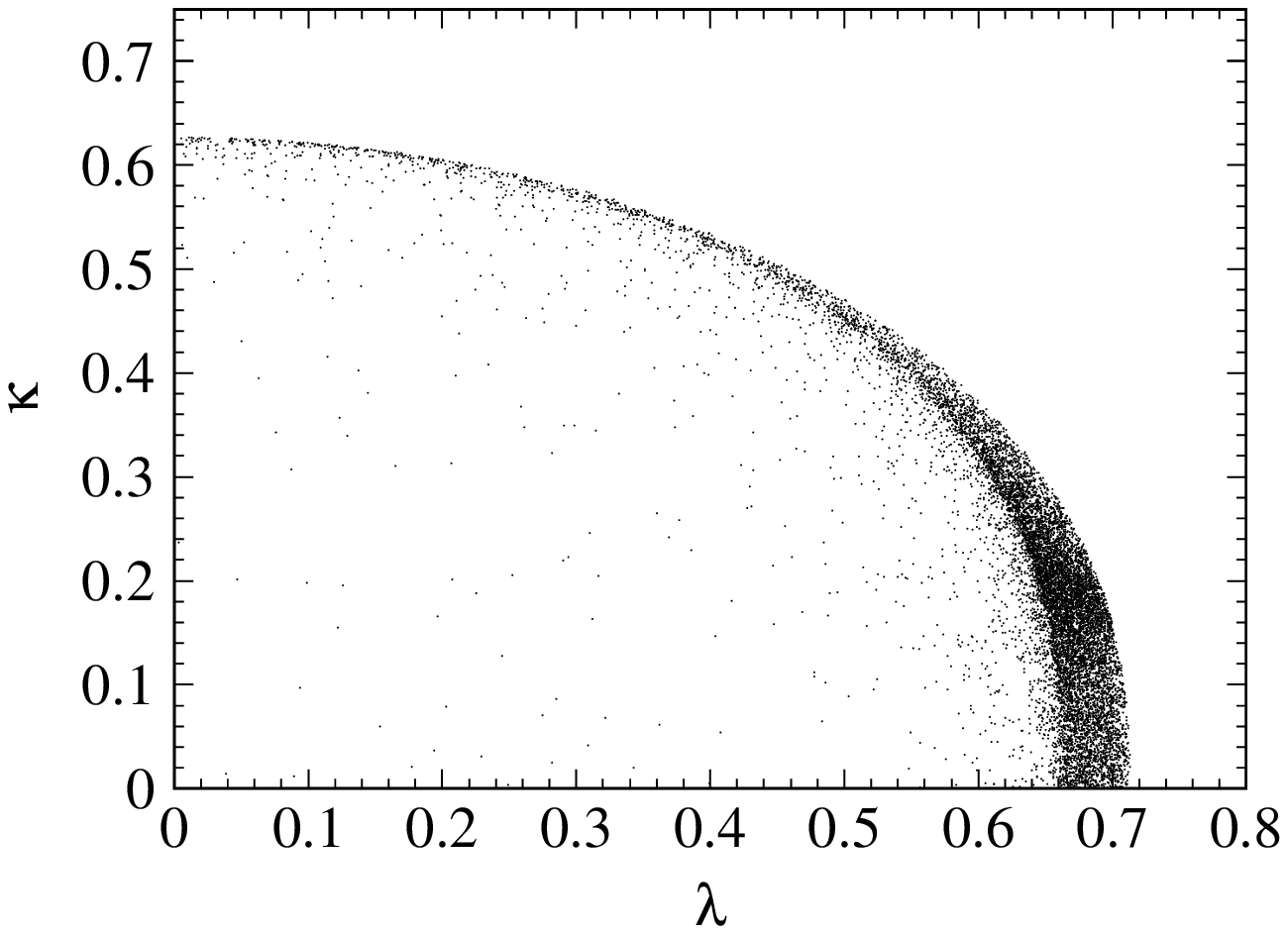}}
\caption{\it Left: The dependence of $\sqrt{\lambda^2+\kappa^2}$
         on renormalization scale, $Q$, for the top Yukawa coupling
         $h_t(M_{\rm GUT})=0.8$. The different curves represent
         different values of $\lambda$ and $\kappa$ at the GUT scale:
         $\lambda=\kappa=1$, $2$, and $3$. {\it Right:} The
         electroweak scale values of $\kappa$ and $\lambda$ for $2
         \times 10^5$ different scenarios with random GUT scale values
         of $0<(\lambda,\kappa,h_t)<2 \pi$. Only scenarios where the
         running top quark mass $m_t(Q)$ falls in the bracket $165 \pm
         5$~GeV at the electroweak scale are retained.  Each point
         represents a different GUT scale parameter choice.}
\label{fig:klscan} 
\end{figure}

Large values of $\lambda$ and/or $\kappa$ at the GUT scale are greatly
reduced when run down to the electroweak scale. This behaviour is
caused by the dependence on $\lambda^4$ and $\kappa^4$ on the
right-hand side of Eqns.(\ref{eq:rgel}--\ref{eq:rgek}) respectively,
indicating that large values of these parameters will evolve strongly,
while small values evolve only slightly. This can be seen in
Fig.(\ref{fig:lkrun}), which shows the dependence of $\lambda$ and
$\kappa$ on renormalization scale. Values of $\lambda$ and $\kappa$ in
the perturbative regime at the GUT scale, i.e.~$\lambda$, $\kappa
\lesssim 2 \pi$, are uniformly reduced to small values at the electroweak scale,
which may be combined to give the approximate bound,
c.f.~Fig.(\ref{fig:klscan}/left),
\be \lambda^2+\kappa^2 \lesssim 0.5. \label{eq:l2k2l}\ee
Also notice the factor of $6$ in Eqn.(\ref{eq:rgek}), compared to the
smaller factors in Eqn.(\ref{eq:rgel}). This tends to make $\kappa$
run more steeply than $\lambda$, which is further exacerbated by the
lack of gauge couplings on the right-hand side of Eqn.(\ref{eq:rgek}),
since the $\kappa$ term in the superpotential contains only the
superfield $\hat S$ which has no gauge couplings.

In order to illustrate these behaviours more clearly, we considered $2
\times 10^5$ different parameter scenarios with (positive) GUT scale
values of $\lambda$, $\kappa$ and $h_t$ chosen randomly (and
independently) between $0$ and $2 \pi$, corresponding to the
perturbative regime defined by $\kappa/2\pi\lesssim 1$, $\lambda/2\pi
\lesssim 1$. Using the renormalization group equations,
Eqns.(\ref{eq:rgeg}--\ref{eq:rgek}), to run $\lambda$ and $\kappa$
down to the electroweak scale, the final distribution in $\lambda$ and
$\kappa$ can be seen in Fig.(\ref{fig:klscan}/right),
where each point is a different parameter choice at the GUT
scale. Only scenarios where the running top quark mass $m_t(Q)$ falls
in the bracket $165 \pm 5$~GeV at the electroweak scale are retained
(approx.~$6\%$).  This plot demonstrates the limit on
$\lambda^2+\kappa^2$ of Eqn.(\ref{eq:l2k2l}) since no parameter point
strays above $\lambda^2+\kappa^2 \approx 0.6$. Furthermore, one can
easily see that most of the parameter points lie at low
$\kappa$. Nevertheless, the entire area within $\lambda^2+\kappa^2
\lesssim 0.6$ is valuable; the accumulation at the boundary results
from the mapping of $\kappa$ and $\lambda$ from the large number of
moderately high values at the GUT scale, to small values at the
electroweak scale.  If $\kappa$ is small at the GUT scale then its
renormalization group running will be weak and it will remain small at
the electroweak scale.

The size of $\kappa$ governs how strongly the U(1) PQ symmetry of the
model is broken. From the above investigation, it is natural to expect
that small values of $\kappa$ are preferred and this symmetry is only
`slightly' broken. However, this is not a strict bound; scenarios with
$\kappa \sim 0.6$ can still occur.

However, if one wants the NMSSM to be a solution to the $\mu$-problem,
$\lambda$ must not become too small. We have introduced $v_s$ as the
VEV of a new Higgs field in order to link $\mu$ to the electroweak (or
supersymmetry) scale. If $\lambda$ becomes too small then the
phenomenological constraints on $\mu$ require $v_s$ to be large and
the link with the other Higgs VEVs is lost. Even if we allow $v_s$ as
large as $3$ TeV (which is a natural upper bound for a weak scale) we
would be forced to maintain $\lambda \gtrsim 0.04$. Consequently the
allowed values of $\lambda$ are rather constrained.

Even more stringent correlations among the parameters are imposed by
grand unified scenarios of the NMSSM including universal boundary
conditions for the soft masses and the trilinear couplings. Radiative
electroweak symmetry breaking requires the couplings, at the
unification scale, $\kappa^2$ to be less than about $\lambda^2$, and
both less than about $0.1$~\cite{ERdTS93}. To prevent QED and color
breaking vacua, the ratio of the universal trilinear coupling and the
scalar mass parameter is restricted to values close to
$3$~\cite{KW95}. If universal boundary conditions are not imposed,
these constraints will not be effective, of course.\\

\underline{$v_s$ and $\tan \beta$} \\[-0.4cm]

The Higgs-higgsino mass parameter $\mu$ is generally assumed to be in
the range $\sim 10^2$--$10^3$~GeV. However, the `natural' scale for
the heavy Higgs masses $M_A$ is seen by Eqn.(\ref{eq:malim}) to be of
the order of $\mu \tan \beta$. Indeed, this is approximately the value
of $M_A$ required for the lightest scalar to reach its maximum
mass. In order to prevent the mass splitting between the light and
heavy Higgs bosons from becoming too large\footnote{Indeed, if a
fairly large splitting were to be observed in experiment, then the
NMSSM with large $\mu$ and/or $\tan \beta$ would provide a very
natural explanation.}, the value of $\tan \beta$ should be kept
moderate, $\lesssim 10$.  The value of $v_s$ could naturally be
expected to vary from $3\,v$ to $15\,v$.

A similar analysis of the renormalization group equations for the
other parameters shows that a low value of $\tan \beta$ is
favoured~\cite{quasi_fixed_point}. Note that the bounds on $\tan
\beta$ derived from the LEP experiments cannot be applied here since
these experiments assume MSSM couplings. Referring to the literature
for the large $\tan \beta$ case~\cite{large_tanb}, here we will adopt a
value of $\tan \beta=3$.\\

\underline{$A_{\kappa}$ and $M_A$} \\[-0.4cm]

\begin{figure}[h]
\bc \includegraphics[scale=0.6]{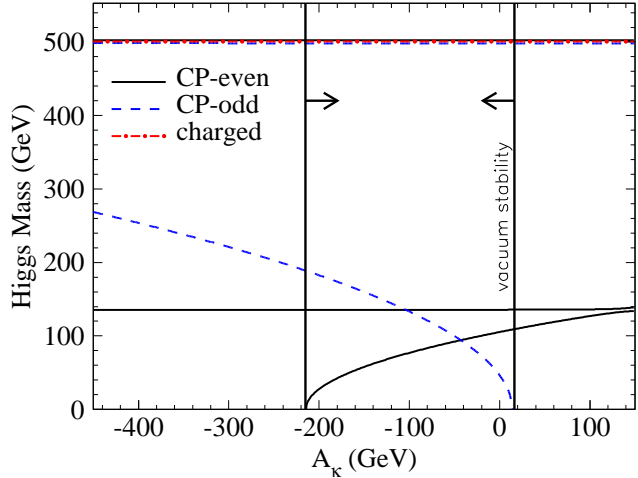} \ec \vspace*{-0.5cm}
\caption{\it The one-loop Higgs boson masses as a function of $A_{\kappa}$ for 
         $\lambda=0.3$, $\kappa=0.1$, $v_s=3\,v$, $\tan \beta=3$ and
         $M_A=\mu \tan \beta \approx 470$ GeV. The arrows denote the
         physically allowed region.}
\label{fig:aklim}
\end{figure}

The value of $A_{\kappa}$ is tightly constrained by the requirement
for vacuum stability, as described in Sec.(2.2).  Note that it occurs
in only one term of each of the CP--even and CP--odd mass matrices,
Eqns.(\ref{eq:m33even}, \ref{eq:m22odd}), where the two contributions
have opposite sign. If $A_{\kappa}$ becomes too large and positive,
it will pull the mass--squared of the lightest pseudoscalar below
zero, destabilizing the vacuum. On the other hand, if it becomes too
large and negative it will destabilize the vacuum by pulling the
mass-squared of the lightest scalar negative. This has been quantified
in Eqn.(\ref{eq:aklim}) and can be seen in Fig.(\ref{fig:aklim})
where the Higgs masses are plotted as a function of $A_{\kappa}$ for
reasonable values of the other parameters, including radiative
corrections.  The heavy Higgs bosons and one of the lighter scalars
are very insensitive to the choice of $A_{\kappa}$, but the
predominantly singlet scalar and pseudoscalar Higgs bosons exhibit a
strong dependence. This is naturally explained by noting that the soft
SUSY breaking term containing $A_{\kappa}$ is a cubic coupling
proportional to $S^3$, so the affect of varying $A_{\kappa}$ is only
communicated through the singlet contribution of the Higgs
fields. Since the mixing is small, this only affects the two fields
which are still singlet dominated.

In contrast to the MSSM, $M_A$ is also constrained by vacuum stability
in the NMSSM, and, as discussed above and visualized in later plots,
its ``natural'' value is approximately $\mu \tan \beta$. Nevertheless,
we will allow $M_A$ to vary over the small bracket of its allowed
range, including the ``natural'' value, and plot the Higgs masses as a
function of $M_A$.

\subsection{The NMSSM with the Peccei--Quinn U(1) Symmetry}

For illustrative purposes, first consider the simplified case where
the U(1) PQ symmetry is left unbroken in the superpotential,
achieved by setting $\kappa=0$. The U(1) symmetry is then
spontaneously broken by the vacuum, giving rise to a massless
Goldstone boson, the PQ axion, which is manifest as the
extra pseudoscalar Higgs field.  Since there is no SUSY breaking term
proportional to $A_{\kappa}$, there are only 4 parameters in the model
at tree--level: $\lambda$, $v_s$, $\tan \beta$ and $M_A$.

While the \underline{charged} Higgs boson masses are, as usual, given
by \vspace*{0.3cm} \\ $(I):$
\vspace*{-0.6cm}
\be M_{H^{\pm}}^2 = M_A^2 +M_W^2- \frac{1}{2} (\lambda v)^2, 
\ee
the tree-level \underline{pseudoscalar} Higgs boson masses read, \vspace*{0.3cm} \\ $(II):$
\vspace*{-0.95cm}
\ba
M_{A_2}^2 &=& M_A^2 \,(1+\frac{1}{4} \cot^2 \beta_s \sin^2 2\beta),\\
M_{A_1}^2 &=&0,
\ea
where the expressions are exact, with no need of approximation.

The approximate solution, where $1/M_A$ and $1/\tan \beta$ are
regarded as small parameters of \cal{O}($\varepsilon$), may again be
used to shed light onto the behaviour of the CP--even masses as the
parameters are varied. However, due to accidental cancellations, the
approximate solution taken to the same order as
Eqn.(\ref{eq:approx12}), \cal{O}($\varepsilon^2$), always yields a
lightest scalar mass-squared that is less than zero. To circumvent
this, one must improve the approximate solution to order
\cal{O}($\varepsilon^4$). For the \underline{scalar} Higgs masses, we find \vspace*{0.3cm} \\ $(III):$
\vspace*{-0.95cm}
\ba
\phantom{I):} M_{H_3}^2 &\approx& M_A^2\, (1+\frac{1}{4} \cot^2 \beta_s \cos^2 2\beta
\sin^2 2\beta) + (M_Z^2-\lambda^2 v^2) \sin^22\beta, \\
M_{H_{2}}^2 &\approx& \frac{1}{2} \left[ M_Z^2
+ \sqrt{ M_Z^4
+\cot^2 \beta_s \left[2\lambda^2v_s^2-M_A^2\sin^22\beta \right]^2}
\right] - (M_Z^2-\frac{1}{2} \lambda^2 v^2)\sin^22\beta, \\
M_{H_{1}}^2 &\approx& \frac{1}{2} \left[ M_Z^2
- \sqrt{ M_Z^4
+\cot^2 \beta_s \left[2\lambda^2v_s^2-M_A^2\sin^22\beta \right]^2}
\right] + \frac{1}{2} \lambda^2 v^2 \sin^22\beta,
\label{eq:approx12k0}
\ea
where in the last term of each, i.e.~the terms of
\cal{O}($\varepsilon^4$), we have made the replacement \mbox{$M_A = \sqrt{2}\lambda
v_s /\sin 2\beta$} for simplicity of the expressions. These relations
improve on those of Ref.\cite{kill_tadpoles2} by describing the
characteristics of the Higgs mass spectrum away from the mid point of
Eqn.(\ref{eq:malim}).

\begin{figure}[ht]
\bc \includegraphics[scale=0.6]{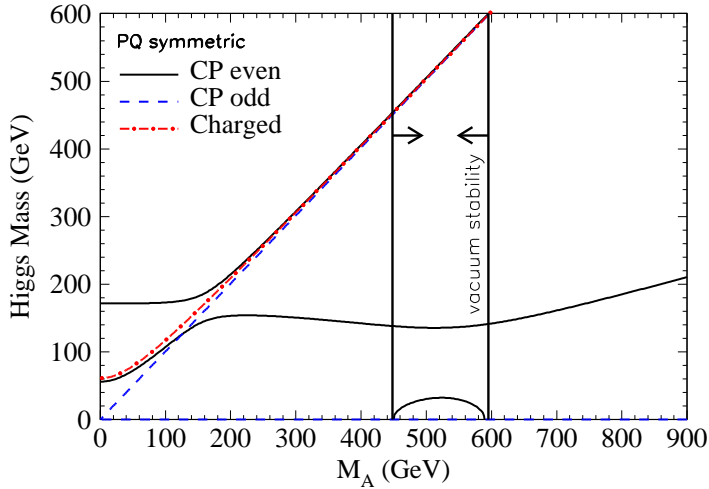} \ec
\caption{\it The one-loop Higgs boson masses, plotted as a function of $M_A$
         for $\lambda=0.3$, $\kappa=0$, $v_s=3v$ and $\tan \beta=3$.
         The arrows denote the physically allowed region.}
\label{fig:masses_scen1}
\end{figure}

The one-loop Higgs boson masses are plotted as a function of $M_A$
with $\lambda=0.3$, $\kappa=0$, $v_s=3v$ and $\tan \beta=3$ in
Fig.(\ref{fig:masses_scen1}), including radiative corrections.
The five heavy Higgs bosons, $H_3$, $H_2$, $A_2$ and $H^{\pm}$, show a
pattern very similar to the MSSM. However the spectrum is augmented by
an additional massless pseudoscalar Goldstone boson and a very light
scalar Higgs boson, conforming with the tree-level upper bound,
$\lambda v \sin 2\beta/\sqrt{2}$, as evident from
Eqn.(\ref{eq:approx12k0}). The positivity of the mass-squared of this
additional scalar Higgs boson places a very stringent bound on the
value of $M_A$, predicting a small bracket in analogy to
Eqn.(\ref{eq:malim}) for the $\kappa$ non-zero case, and consequently
strongly constraining the masses of the Higgs bosons $H_3$, $A_2$ and
$H^{\pm}$.

Taking $\lambda \to 0$, besides the massless pseudoscalar PQ axion,
also the mass of the additional scalar Higgs boson approaches
zero. Both fields decouple from the doublet Higgs sector. At the same
time, however, the degenerate masses of the heavy Higgs bosons are
fixed to the approximate value $M_A \approx \mu \tan \beta$. For
$\kappa=0$, the ``MSSM limit'' is approached on a trajectory $\lambda
\to 0$ that is constrained to a sheet in the MSSM parameter space that
respects the PQ symmetry in the final MSSM potential. In fact, the
$H_uH_d$ mixing term, that violates the PQ symmetry in the Higgs
potential, is suppressed compared to the diagonal $|H_u|^2$ term,
which respects the PQ symmetry, in the limit of large $M_A$ and $\tan
\beta$ analyzed analytically above.

\begin{figure}[ht]
\hspace*{-1cm}\includegraphics[scale=0.65]{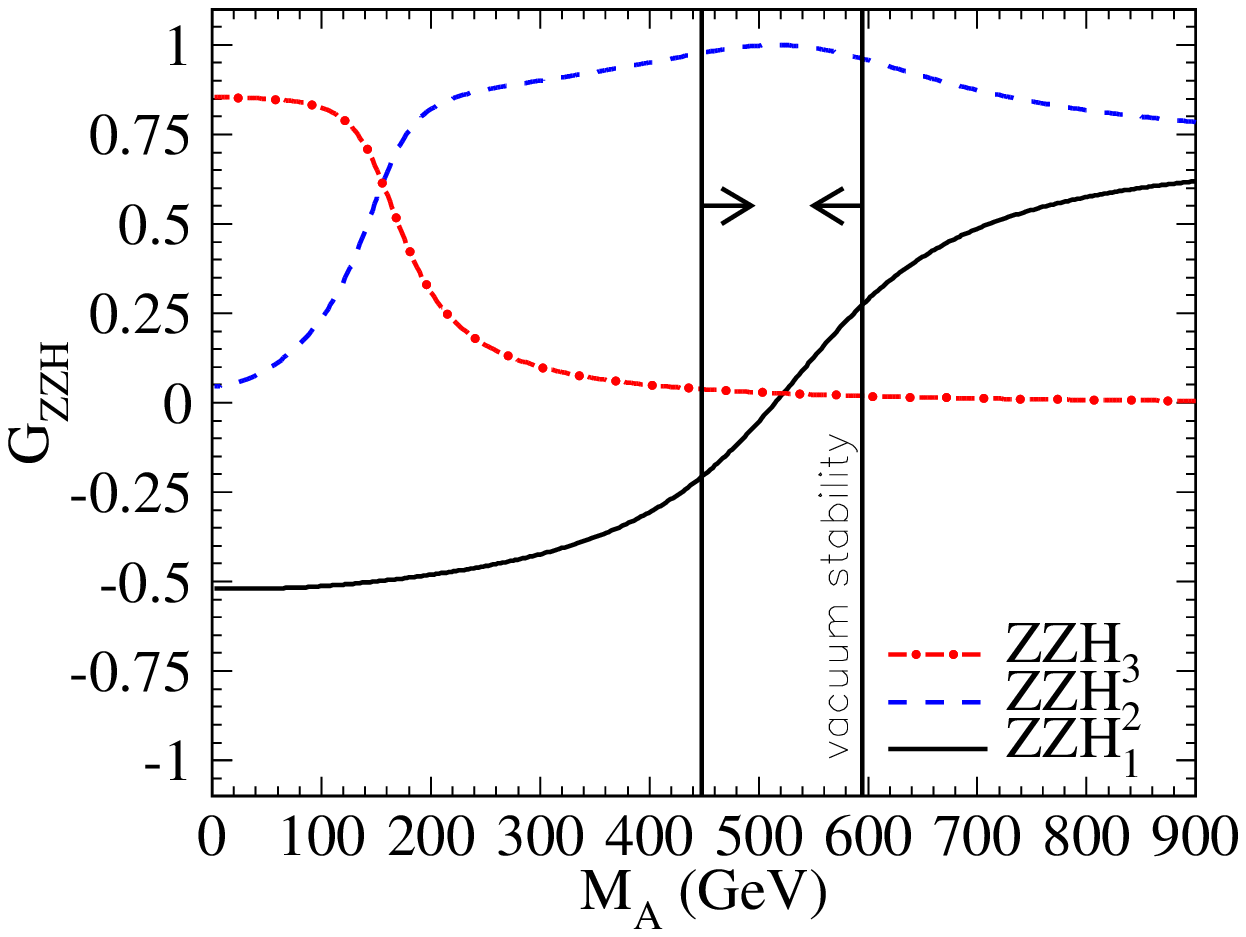}\hspace*{-1cm}
\includegraphics[scale=0.65]{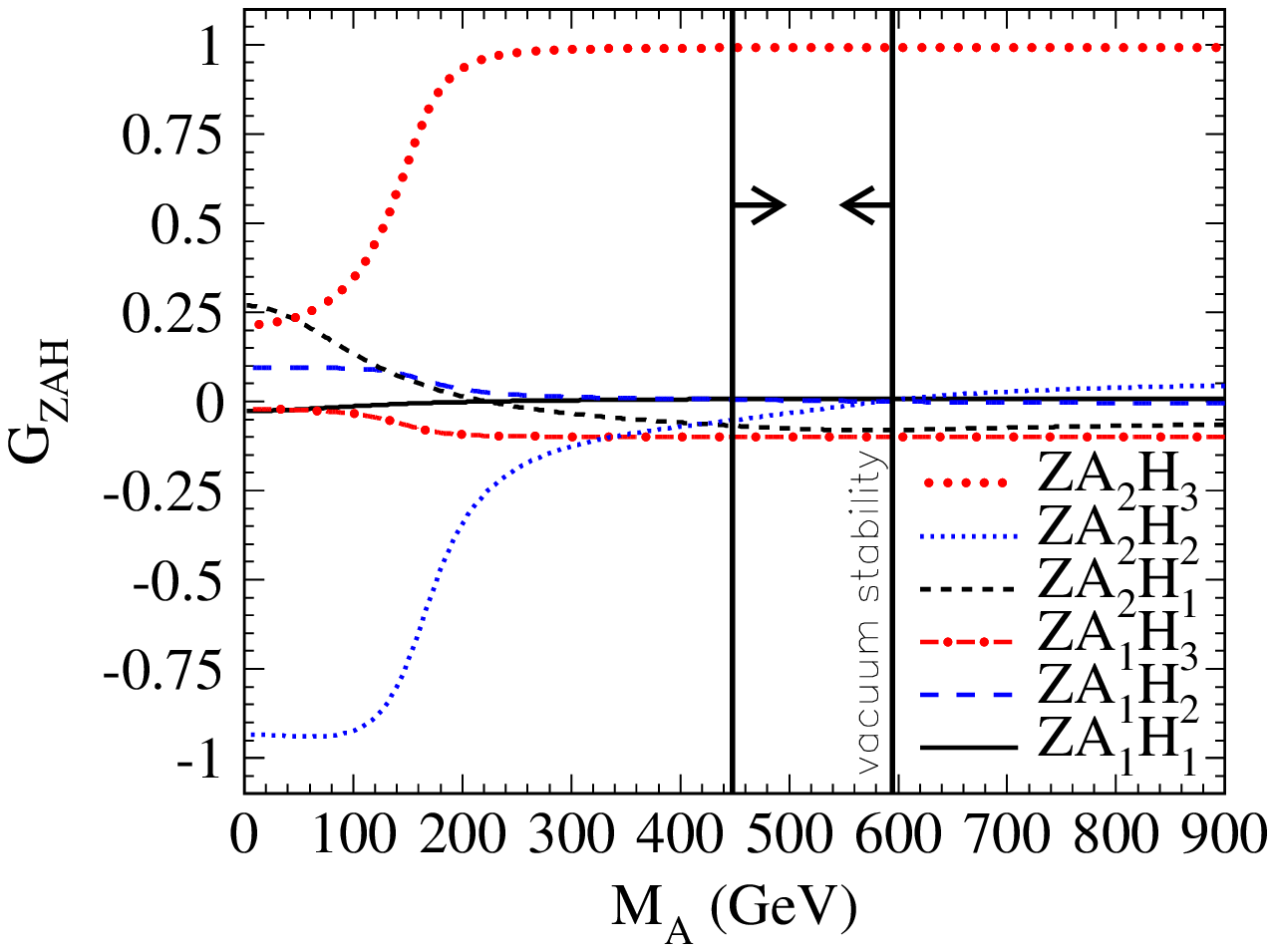}
\caption{\it The couplings $\mathcal{G}_{ZZH_i}$ (left) and $\mathcal{G}_{ZA_iH_j}$ (right),
         plotted as a function of $M_A$ for $\lambda=0.3$, $\kappa=0$,
         $v_s=3v$ and $\tan \beta=3$.  The arrows denote the
         physically allowed region.}
\label{fig:coup_scen1}
\end{figure}

While the present model may serve as a nice illustration for the
complex phenomena in the NMSSM, the presence of a massless
pseudoscalar and light scalar rules out this parameter choice, due to
their couplings to the $Z$ boson. The direct production of the
lightest scalar together with a $Z$ boson, $Z \to Z H_1$, the coupling
of which is shown in Fig.(\ref{fig:coup_scen1}/left),
would have been possible at LEP for much of the $M_A$ range. However,
since the coupling passes through zero, such production alone does not
yet rule out the model. The $Z$ boson decay $Z \to H_1 A_1$ is more
damaging, since the coupling $\mathcal{G}_{ZA_1H_1}$ (see
Fig.(\ref{fig:coup_scen1}/right)), although small, is always large
enough to allow detection.

This may only be remedied by forcing $\lambda$ to become very small,
so that the extra scalar and pseudoscalar fields decouple from the SM
gauge bosons. Disregarding potential fine-tuning problems, these
models are further constrained by astrophysical axion searches and
cosmological bounds which are only satisfied for $10^{-10} < \lambda
<10^{-7}$~\cite{axion_constraints}. Since $\mu$ is constrained to be
of the order of the electroweak scale, this requires the extra singlet
field to have a very large VEV, making this model unattractive as a
solution to the $\mu$-problem.

\subsection{The NMSSM with a slightly broken Peccei--Quinn symmetry}

Turning on a non-zero value of $\kappa$ breaks the PQ $U(1)$ symmetry,
providing a non-zero mass for the lightest pseudoscalar Higgs boson,
and raising the mass of the lightest scalar Higgs boson. We consider
the PQ symmetry to be only `slightly' broken as long as $\kappa \ll
1$, for values of $v_s$ of the order of the electroweak scale. This is
favoured by the renormalization group flow as illustrated in our
earlier discussion. The model contains two extra parameters as
compared to the case with an unbroken PQ symmetry: $\kappa$ and its
associated soft SUSY breaking parameter $A_{\kappa}$.

As a representative example of this scenario we set $\lambda=0.3$ and
$\kappa=0.1$.  We will choose a value of $A_{\kappa}$ in the centre of
the allowed range, $A_{\kappa}=-100$ GeV, c.f.~Fig.(\ref{fig:aklim}),
but the masses of the two singlet dominated fields can change
significantly by altering this value.

\begin{figure}[h]
\bc \includegraphics[scale=0.6]{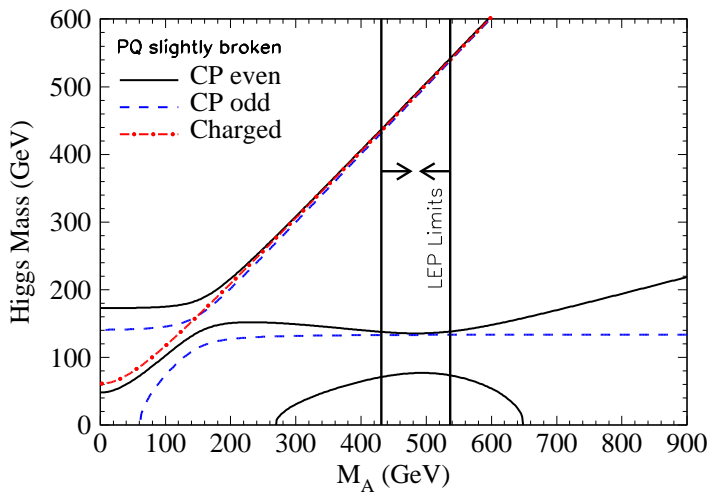} \ec
\caption{\it The one-loop Higgs boson masses as a function of 
$M_A$ for $\lambda=0.3$, $\kappa=0.1$, $v_s=3\,v$, $\tan \beta=3$ and
$A_{\kappa}=-100$ GeV. The arrows denote the region allowed by LEP
searches with 95\% confidence.}
\label{fig:masses_scen2a}
\end{figure}

The masses of the Higgs bosons for $\lambda=0.3$, $\kappa=0.1$, $v_s=3
v$, $\tan \beta=3$ and $A_{\kappa}=-100$ GeV are shown as a function
of $M_A$ in Fig.(\ref{fig:masses_scen2a}), including radiative
corrections.
The structure of the heavy Higgs boson masses, $H_3$, $A_2$ and
$H^{\pm}$, remains very similar to the MSSM. However, the light Higgs
boson mass spectrum is distinctly different, with two light scalar
Higgs bosons and one light pseudoscalar Higgs boson. The extra singlet
dominated pseudoscalar is no longer massless. Its mass has been raised
from zero, by having the PQ symmetry explicitly broken by the term
$\frac{1}{3} \kappa S^3$ in the superpotential, to a value largely
independent of $M_A$, c.f.~Eq.(\ref{eq:ma1app}). Once again, the value
of $M_A$ is bounded by the requirement that the vacuum be stable
(i.e.~$M^2_{H_1}>0$), although the restriction is now looser.  The
$H_2$ mass, of the order of $M_Z$ and largely independent of
$M_A$, can rise above the canonical MSSM value due to the interaction
with the new singlet fields, c.f.~Eq(\ref{eq:m22even}). Generating a
spectrum of light Higgs states, the NMSSM would be easily
distinguished from the MSSM for this parameter range except in small
regions where the $ZZH_i$ couplings are particularly close to zero.

\begin{figure}[h]
\hspace*{-1cm}\includegraphics[scale=0.65]{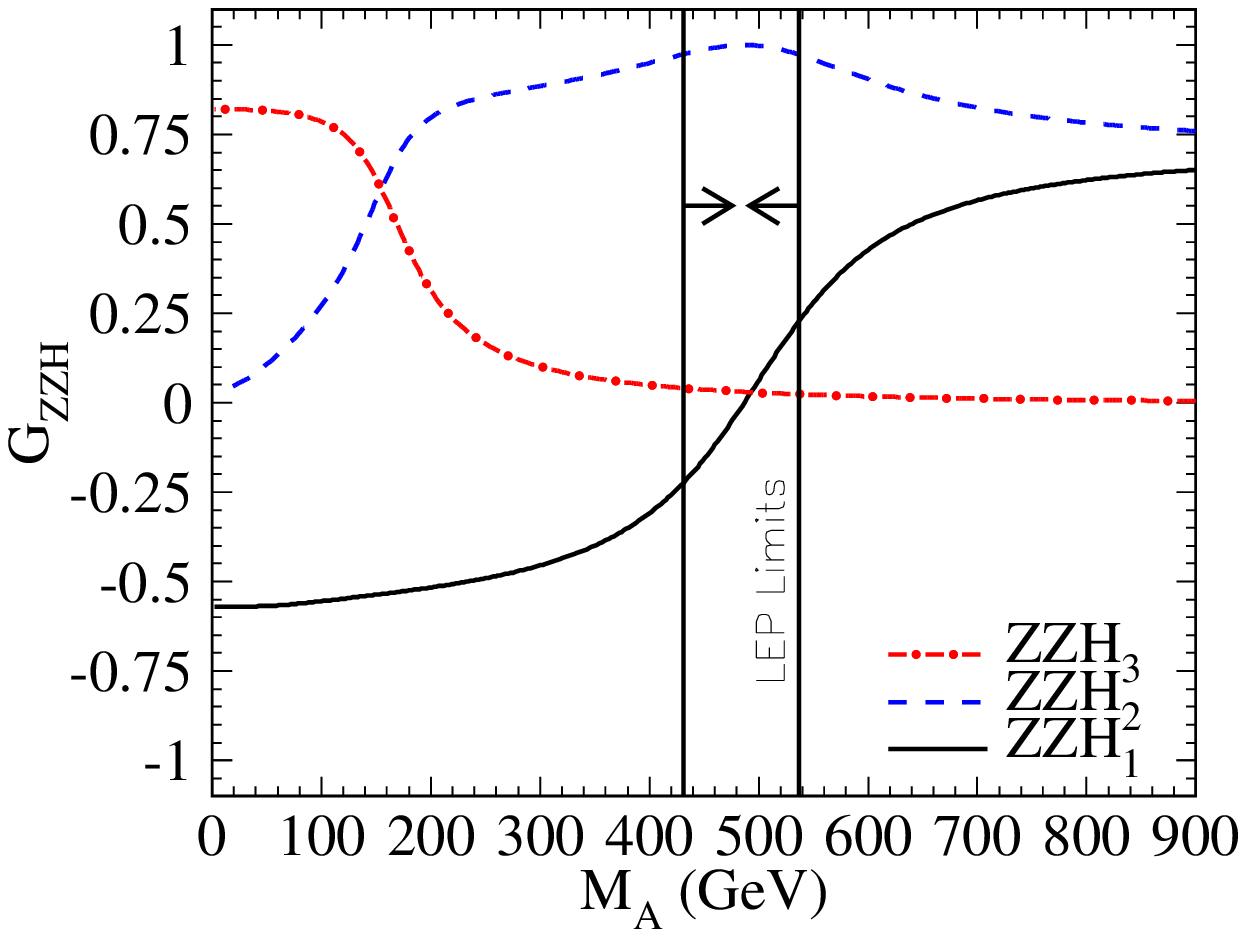}\hspace*{-1cm}
\includegraphics[scale=0.65]{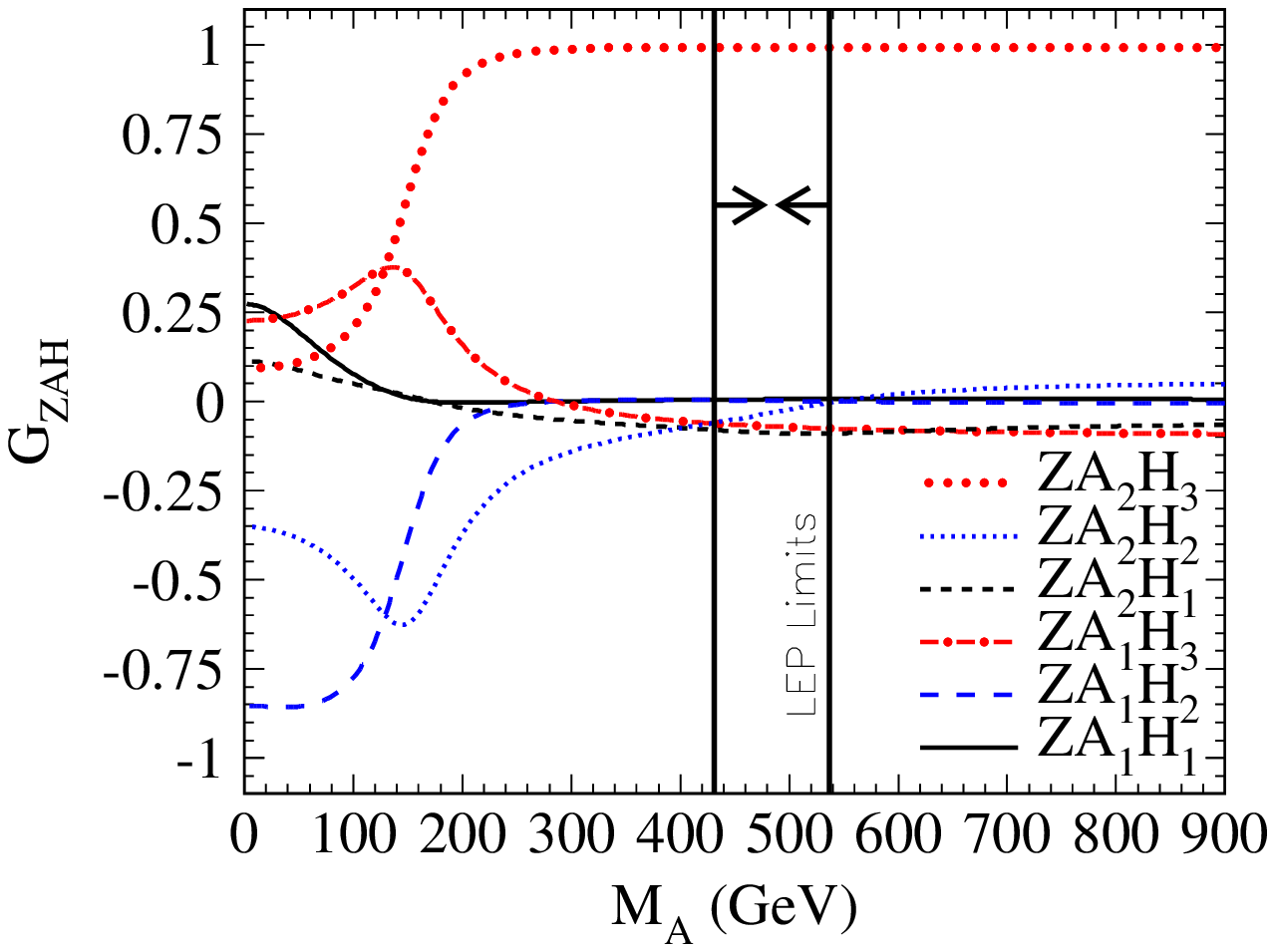}
\caption{\it The couplings $\mathcal{G}_{ZZH_i}$ (left) and $\mathcal{G}_{ZA_iH_j}$ (right),
as a function of $M_A$ for $\lambda=0.3$, $\kappa=0.1$, $v_s=3\,v$,
$\tan \beta=3$ and $A_{\kappa}=-100$ GeV. The arrows denote the region
allowed by LEP searches with 95\% confidence.}
\label{eq:coup_scen2a}
\end{figure}

Also shown are the restrictions on $M_A$ imposed by the LEP
experiments~\cite{LEP_wg} with 95\% confidence. These limits are
dependent on the masses of the light Higgs bosons and their couplings
to the $Z$, as shown in Fig.(\ref{eq:coup_scen2a}).
The light scalar escapes detection via Higgs-strahlung only when it
has a sufficiently reduced coupling to the $Z$ boson. Since this
coupling passes through zero in the allowed range, one is unable to
rule out all values of $M_A$. Furthermore, in contrast to the
$\kappa=0$ scenario, the lightest pseudoscalar is now heavy enough to
escape production via $e^+e^- \to H_1A_1$ at LEP. If the couplings of
the lightest Higgs boson are particularly small, this could be
difficult to discover even at the LHC\footnote{The couplings of the
new fields to quarks and leptons are manifest only from their mixing
with the usual Higgs doublet fields, just as for their couplings to
the gauge bosons. Small mixings will therefore also suppress the
(loop) coupling of the predominantly singlet Higgs bosons to
gluons. Searches for Higgs decays to light axion pairs at colliders
have been discussed in Ref.\cite{axion_pairs}.}. It is therefore
important that a future linear collider, such as
JLC/NLC/TESLA~\cite{JLC}--\cite{TESLA} or CLIC~\cite{CLIC} perform a
search for light Higgs bosons with reduced couplings\footnote{For the
phenomenology of Higgs-strahlung, $e^+e^-
\to ZH_i$, and associated scalar/pseudoscalar production, $e^+e^- \to
H_iA_j$, for heavy and light Higgs bosons, see Ref.\cite{acc}.}.\\

Varying the parameters $v_s$ and $A_{\kappa}$ changes the
quantitative, if not qualitative, features of the mass spectrum. In
particular, large departures from the typical MSSM Higgs mass pattern
occur in the light sector.  As pointed out in Sec.(3.1), changing the
value of $A_{\kappa}$ varies the lightest pseudoscalar and scalar
masses while leaving the heavier Higgs bosons unaffected (see
Fig.~(\ref{fig:aklim})).

\begin{figure}[h]
\hspace*{-1cm}\includegraphics[scale=0.65]{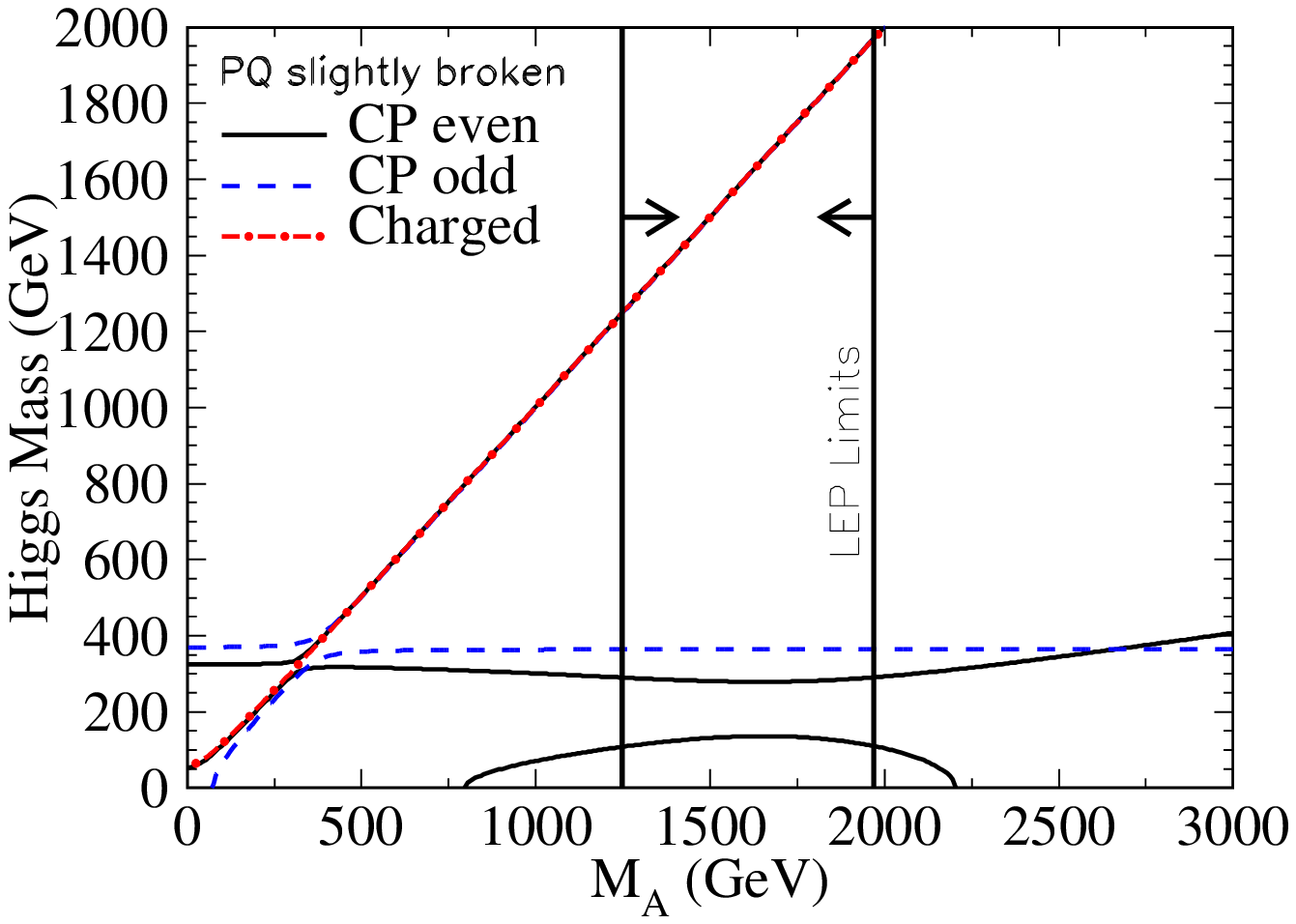}\hspace*{-1cm}
\includegraphics[scale=0.65]{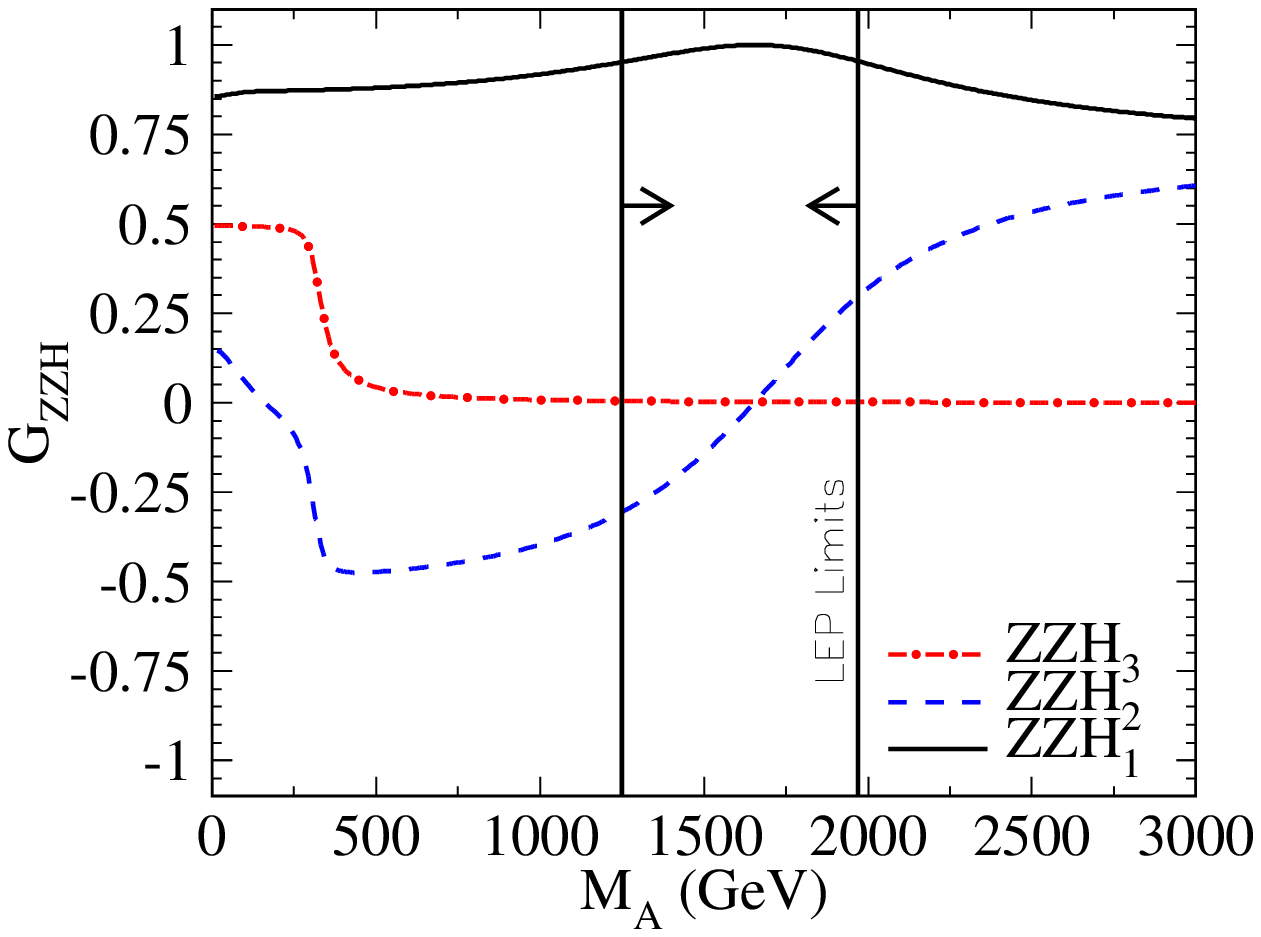}
\caption{\it The one-loop Higgs boson masses(left) and couplings 
$\mathcal{G}_{ZZH_i}$ 
         (right) as a function of $M_A$ for $\lambda=0.3$,
         $\kappa=0.1$, $v_s=10\,v$, $\tan \beta=3$ and
         $A_{\kappa}=-250$ GeV. The arrows denote the region allowed
         by LEP searches with 95\% confidence.}
\label{fig:scen2b}
\end{figure}

If $v_s$ (or $\lambda$) is increased, all but the lightest Higgs boson
become heavier.  This can be seen in Fig.(\ref{fig:scen2b}/left)
where the Higgs boson masses have been plotted with $v_s=10\,v$,
\mbox{$A_{\kappa}=-250$~GeV} and the other parameters as
before\footnote{$A_{\kappa}$ has been chosen to lie in the middle of
its allowed range.}.  The restrictions on $M_A$ enforced by vacuum
stability (i.e.~$M_{H_1}^2>0$) are relaxed, allowing a larger range of
$M_A$ values. One should note that \mbox{$M_A \approx \sqrt{2}\lambda
v_s /\sin 2 \beta \approx \mu \tan \beta$} remains a point of
particular interest as this choice allows the largest mass for
$H_1$. $M_A$ is now forced to be beyond the $1$~TeV range.

Significantly, the increase of $v_s$ causes the mixing to become
stronger, increasing the doublet component in $H_1$ at the expense of
the doublet component in $H_2$. Consequently, the character of $H_1$
becomes more doublet-like, while the character of $H_2$ becomes more
singlet-like. This increased mixing is quantified by
Eqn.(\ref{eq:tantha}). Consequently, the lightest scalar Higgs boson
acquires a significant coupling to the $Z$ boson, as shown in
Fig.(\ref{fig:scen2b}/right).\\

\begin{figure}[h]
\hspace*{-1cm}\includegraphics[scale=0.65]{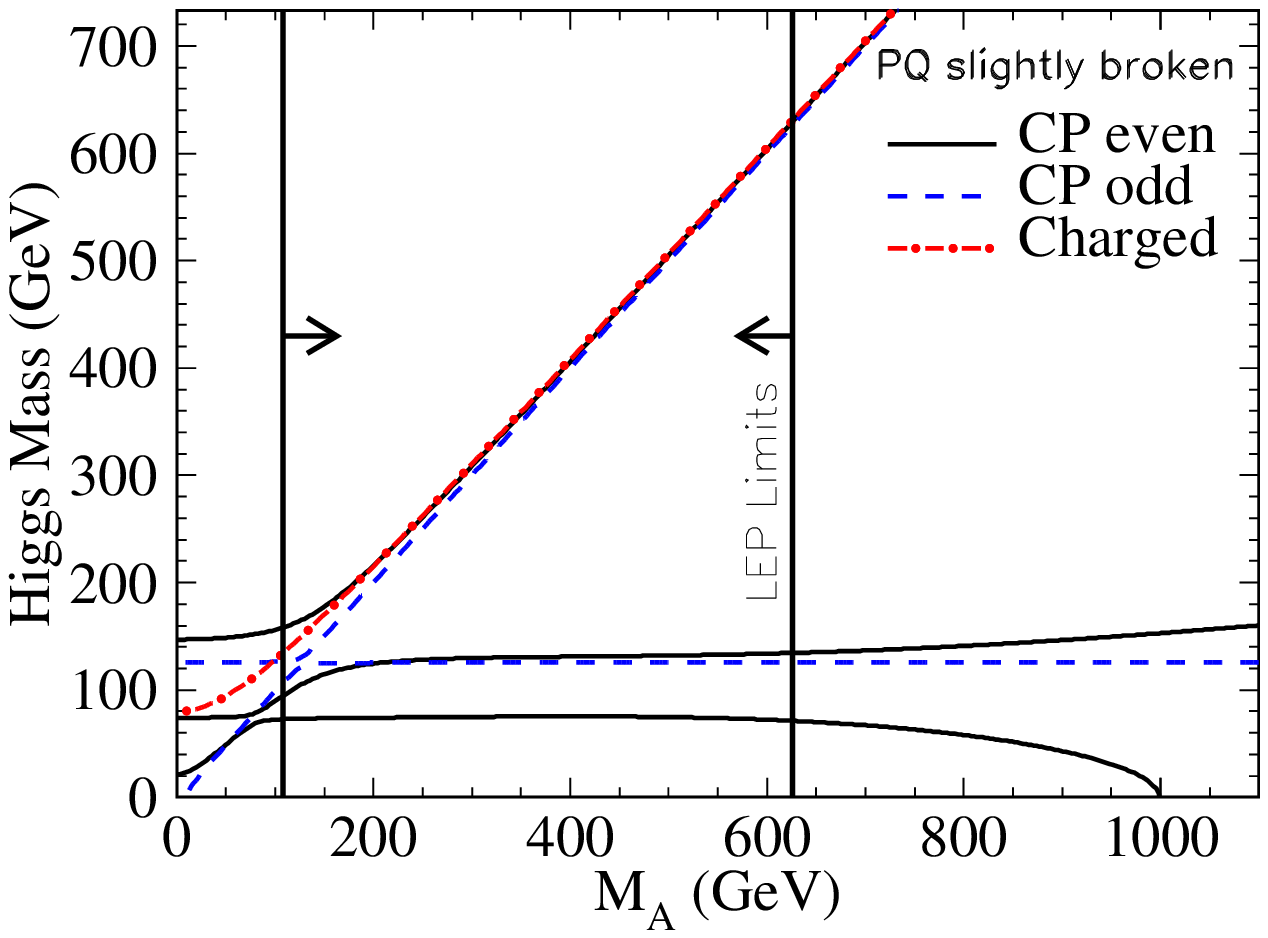}\hspace*{-1cm}
\includegraphics[scale=0.65]{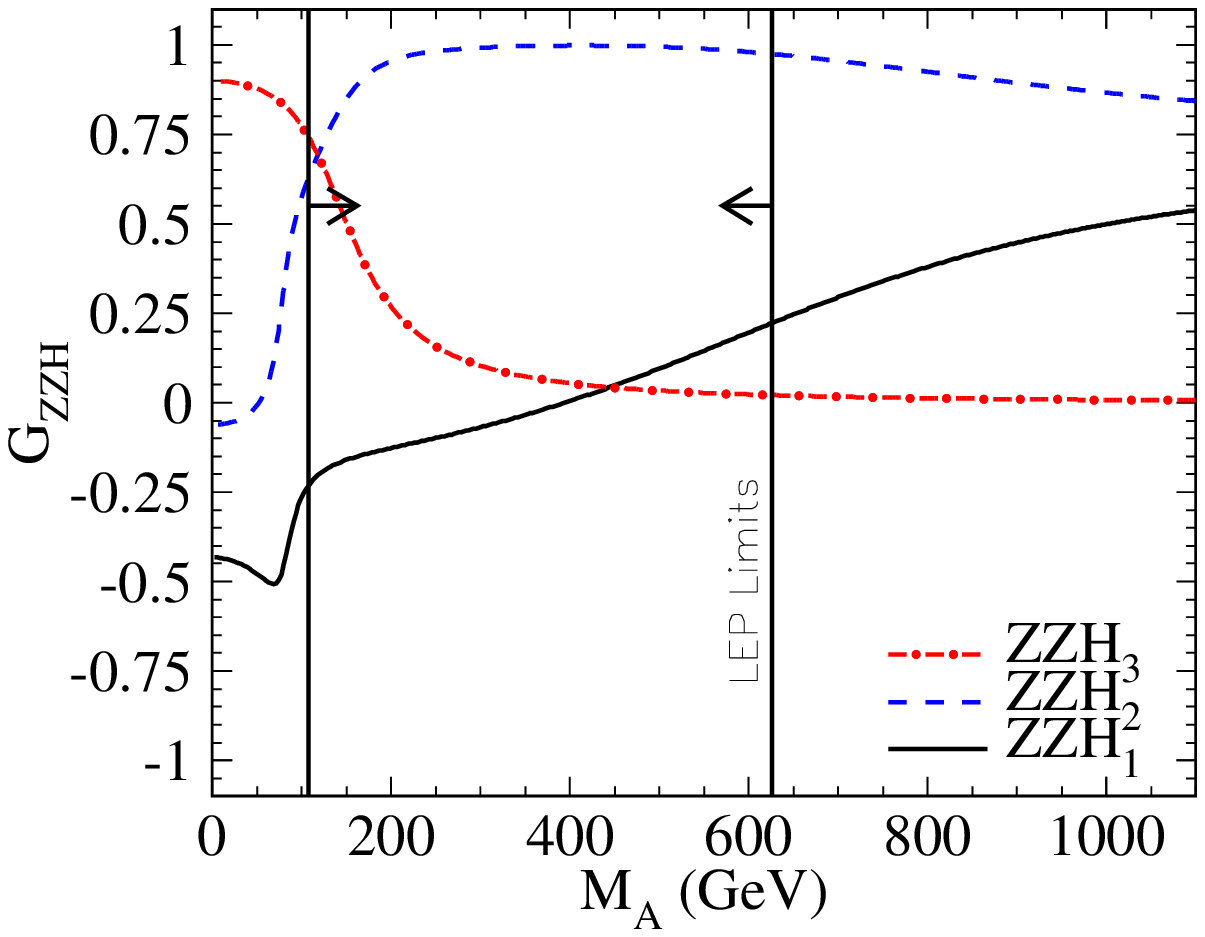}
\caption{\it The one-loop Higgs boson masses (left) and 
couplings $\mathcal{G}_{ZZH_i}$ 
(right) as a function of $M_A$ for $\lambda=0.05$, $\kappa=0.02$,
$v_s=15\, v$, $\tan \beta=3$ and $A_{\kappa}=-100$ GeV. The arrows
denote the region allowed by LEP searches with 95\% confidence.}
\label{fig:scen2c}
\end{figure}

A further example of models in this class is given by $\lambda=0.05$,
$\kappa=0.02$. Such a small value of $\lambda$ requires a large value
of $v_s$, here taken to be $15 \, v$, in order to satisfy the
phenomenological constraints on $\mu$. The Higgs mass spectrum and
couplings for this model remain qualitatively the same, as can be seen
in Fig.(\ref{fig:scen2c}), although a larger range of $M_A$ values
remains viable.
 \\

Finally, changing the sign of $\kappa$, while keeping that of $\kappa
A_{\kappa}$ fixed, results in behaviours very similar to their
positive $\kappa$ counterparts.  Changing the sign of $\kappa
A_{\kappa}$ while keeping that of $\kappa$ fixed results in an
unstable vacuum, as can be seen in Fig.(\ref{fig:aklim}).

\subsection{The NMSSM with a strongly broken Peccei--Quinn symmetry}

\begin{figure}[p]
\bc \includegraphics[scale=0.6]{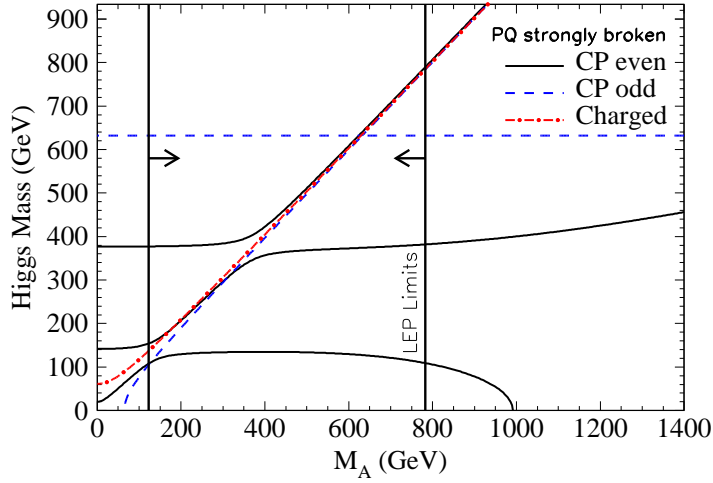} \ec
\caption{\it The one-loop Higgs boson masses as a function of $M_A$ for 
         the parameter choice $\lambda=0.3$, $\kappa=0.5$, $v_s=3\,v$,
         $\tan \beta=3$ and $A_{\kappa}=-500$~GeV. The arrows denote
         the region allowed by LEP searches with 95\% confidence.}
\label{fig:masses_scen3}
\end{figure}

When $\kappa$ becomes large the PQ symmetry is strongly broken. This
scenario has been shown earlier to be disfavoured by the
renormalization group flow but cannot be ruled out {\it a priori} on
general grounds. The parameters $v_s$, $M_A$ {\it etc.\ } are taken to
be of the order of the electroweak symmetry breaking scale.  The extra
pseudoscalar and scalar Higgs bosons gain moderately large masses from
the PQ breaking term $\frac{1}{3} \kappa S^3$ in the superpotential,
with values $M_{A_1}^2 \sim -3 \kappa v_s A_{\kappa}/\sqrt{2}$ and
$M_{H_2}^2 \sim \kappa v_s (\kappa v_s +\sqrt{2}A_{\kappa})/2$,
c.f.~Eqns.(\ref{eq:ma1app}) and (\ref{eq:approx12}), largely
independent of $M_A$.  The eigenstate predominantly composed of the
new scalar may in general no longer be the lightest scalar Higgs
boson. That r\^{o}le is taken over by a Higgs boson which is
predominantly composed of the doublet field $S_2$, with a mass of
order $M_{H_1}\sim M_Z$ and a SM like coupling to the $Z$.  The masses
of these Higgs bosons are shown as a function of $M_A$ for the
parameter choice $\lambda=0.3$, $\kappa=0.5$, $v_s=3
\, v$, $\tan
\beta=3$ and $A_{\kappa}=-500$ GeV, in Fig.(\ref{fig:masses_scen3}).
\begin{figure}[p]
\hspace*{-1cm}\includegraphics[scale=0.65]{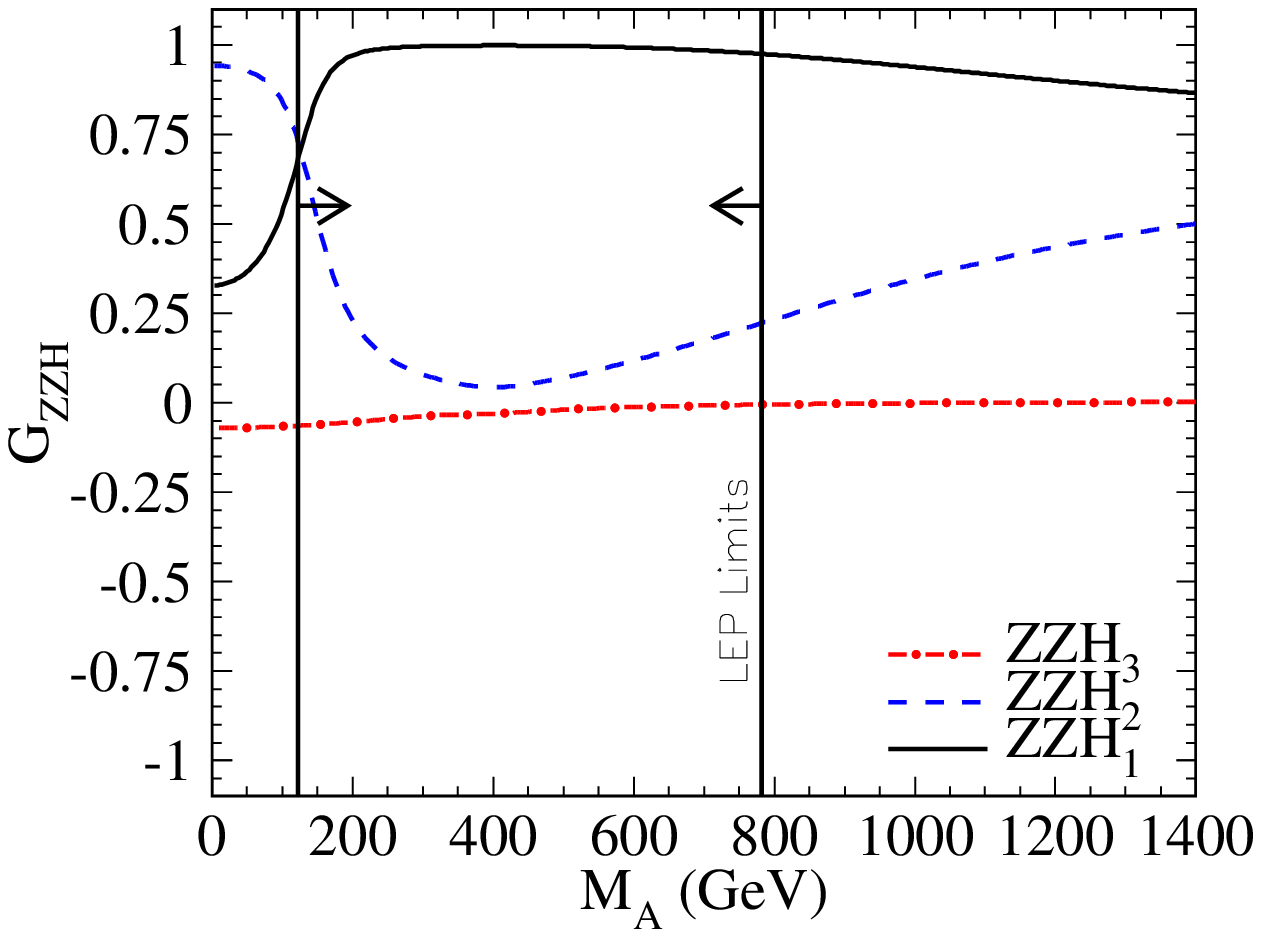}\hspace*{-1cm}
\includegraphics[scale=0.65]{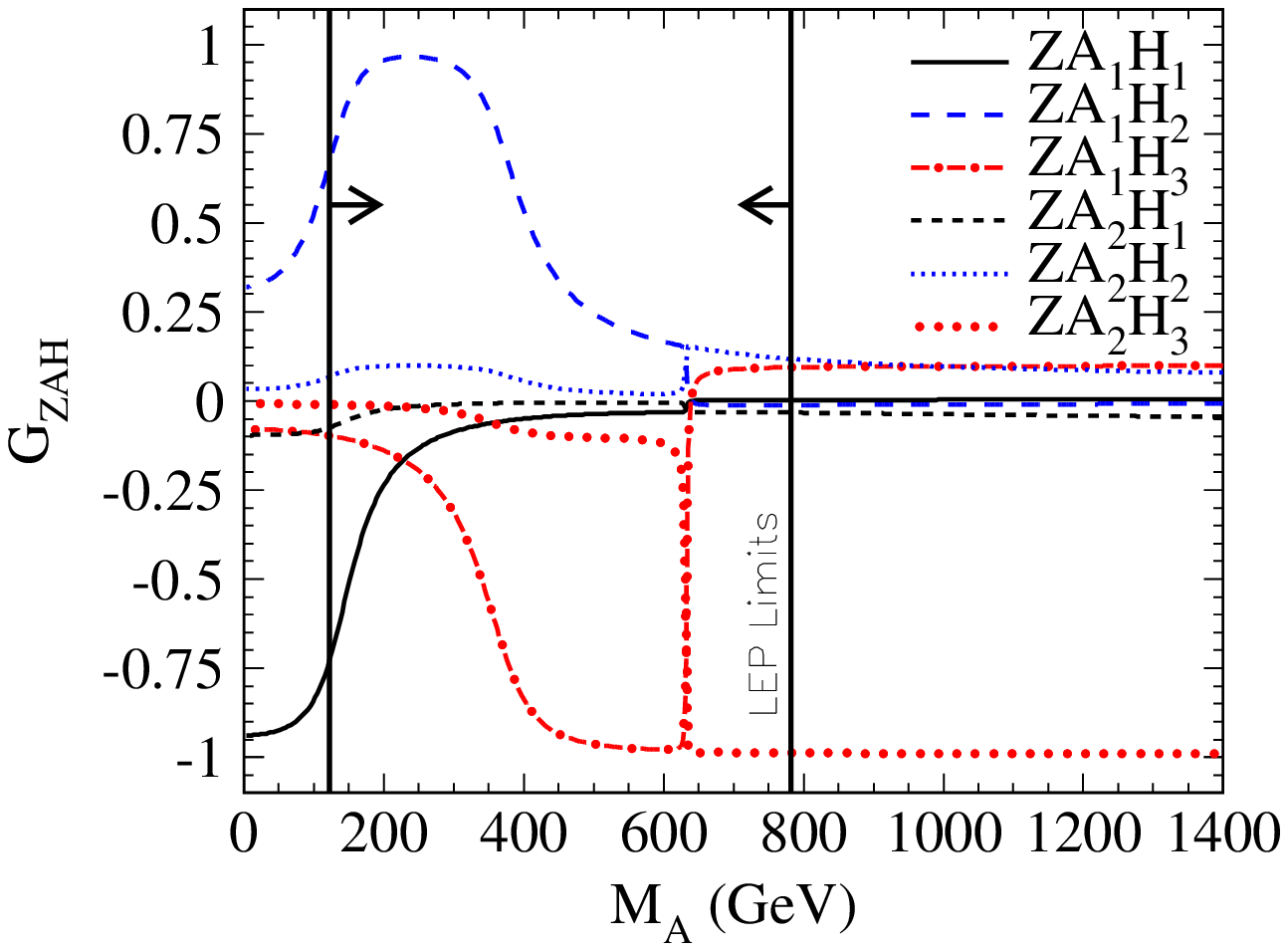}
\caption{\it The couplings $\mathcal{G}_{ZZH_i}$ (left) and $\mathcal{G}_{ZA_iH_j}$ (right),
	as a function of $M_A$ for the parameter choice $\lambda=0.3$,
	$\kappa=0.5$, $v_s=3\,v$, $\tan \beta=3$ and
	$A_{\kappa}=-500$~GeV. The arrows denote the region allowed by
	LEP searches with 95\% confidence.}
\label{fig:coup_scen3}
\end{figure}
Again the predominantly doublet fields follow the pattern of the MSSM,
with one CP-even Higgs boson, $H_1$, below the electroweak scale and
one CP-even, one CP-odd and two charged Higgs bosons nearly degenerate
at the scale $M_A$. However, now the predominantly singlet scalar and
pseudoscalar fields are heavy enough that they are no longer the
lightest states for much of the physically allowed range, being,
independently of $M_A$, at the scale of a few hundred GeV.

The couplings to the $Z$ boson, Fig.(\ref{fig:coup_scen3}), show a
complicated structure, due mainly to the way in which the Higgs states
are labeled. As $M_A$ is increased a previously light Higgs boson may
gain sufficient mass to overtake one of the other Higgs bosons. The
labeling will change (according to the mass hierarchy) causing an
apparent rapid variation of the $Z$ couplings at the crossing. It is
this relabeling which gives the complicated structure of the $ZA_iH_j$
couplings.

Increasing the value of $v_s$ will increase the mass of the new fields
substantially (since the effect of the increase seen in the previous
section is magnified by the large value of $\kappa$).  For example,
with the same parameters as Fig.(\ref{fig:masses_scen3}) except
$v_s=10\,v$, the new scalar and pseudoscalar states have masses of the
order of $1.6$ TeV and $1.2$ TeV, respectively, for all values of $M_A$
(although these masses show a strong dependence on $A_{\kappa}$). The
new fields will also decouple, making the NMSSM very difficult to
distinguish from the MSSM.

Once again, negative values of $\kappa$ do not change the qualitative
picture as long as one also changes the sign of $A_{\kappa}$.

\section{Summary and Conclusions}

In this study, we have investigated the Higgs sector of the
Next-to-Minimal Supersymmetric Standard Model, suggested by many GUT
and superstring models. Moreover, this model attempts to explain the
$\mu$-problem of the MSSM by introducing a new singlet Higgs field,
$S$, with a non-zero vacuum expectation value. When this new field is
coupled to the Higgs doublets, its expectation value leads to the
$\mu$ term of the MSSM, providing a value of $\mu$ which may naturally
be expected near the electroweak scale.

We have given expressions for the Higgs boson mass matrices and
presented, besides the numerical analyses, approximate analytical
solutions for the charged, CP-even and CP-odd Higgs boson masses which
provide a nice insight into the mass hierarchies.  It has been found
that useful sum rules and inequalities for the mass parameters can be
established and that the requirement of vacuum stability can place
useful bounds on the mass parameters and couplings of the Lagrangian.

The renormalization group flow of the parameters $\lambda$ and
$\kappa$ from the GUT scale down to the electroweak scale provide
strong upper bounds on their values at the electroweak scale, where
small $\kappa$ is favoured.  The qualitative features of the Higgs
boson masses are dependent on how strongly the PQ symmetry of the
model is broken, quite accurately described by the approximate
analytical solutions.

If the PQ symmetry is left explicitly unbroken in the Lagrangian
(spontaneously broken by the structure of the vacuum), a massless
Goldstone boson is present, the PQ axion. Additionally, this model
contains two CP-even Higgs bosons with masses of the order of the
electroweak scale or lower. As in the MSSM, the heavy fields, one
CP-odd, one CP-even and two charged Higgs bosons, are nearly
degenerate, Fig.(\ref{fig:masses_scen1}). However, unlike the MSSM,
the vacuum structure of the model constrains these heavy states to lie
close in mass to $\mu \tan \beta$.  Non-observation of the PQ axion
rules out most of the allowed parameter space, only allowing scenarios
with very low values of $\lambda$ and therefore very high VEVs for the
new singlet field, which thereby do not provide a natural solution of
the $\mu$-problem.

If the PQ symmetry is slightly broken, the qualitative pattern for the
particle spectrum remains intact, except that the lightest CP-odd
state acquires a mass of the order of the electroweak scale.  Thus the
model contains a heavy CP-even and CP-odd Higgs boson plus two heavy
charged Higgs bosons which are all nearly degenerate in mass,
similarly to the MSSM. However, besides the standard light CP-even
Higgs boson, the spectrum is complimented by a pair of CP-even and
CP-odd Higgs bosons with masses below the electroweak symmetry
breaking scale, Fig.(\ref{fig:masses_scen2a}).  The LEP limits on the
lightest Higgs boson mass allow some of the parameter space to be
ruled out. However, since the couplings to the $Z$ boson can be very
much reduced, the NMSSM with a slightly broken PQ symmetry constitutes
a valid scenario. Observing three light Higgs bosons, but no charged
light Higgs boson, at future colliders would present an opportunity to
distinguish the NMSSM with a softly broken PQ symmetry from the MSSM,
even if the heavy states are inaccessible. It is therefore important
that future colliders search for light scalar and pseudoscalar Higgs
bosons with reduced couplings.

In contrast, a strongly broken PQ symmetry, though disfavoured by the
flow of the couplings from the GUT scale down to the electroweak
scale, could provide extra moderately heavy Higgs bosons,
Fig.(\ref{fig:masses_scen3}), which are only weakly coupled to the $Z$
boson. Such decoupled scenarios would be more difficult to distinguish
from the MSSM.\\[0.2cm]

{\it Acknowledgments:} The authors are grateful to S.~Dedes,
C.~Hugonie, S.~Moretti, H.B.~Nielsen and V.~Rubakov for fruitful
discussions. RN is indebted to Alfred T\"opfer Stiftung and to the
DESY Laboratory for a scholarship during his stay in Hamburg
(2001--2002). The work of RN was also supported by the Russian
Foundation for Basic Research (RFBR), projects 00-15-96562 and
02-02-17379.

\vspace{1cm}


\section*{Appendix: An approximate solution}

The CP--even mass matrix of Eqns.(\ref{eq:m11even}--\ref{eq:m33even})
does not lend itself easily to obtaining analytic expressions for the
physical Higgs masses. However, reasonably simple expressions can be
found by making an approximation, providing expressions which may be
used to shed some light on the behaviour of the Higgs masses as the
other parameters are varied.

To construct this approximate solution in the scalar sector we regard
both $1/\tan \beta$ and $1/M_A$ as small parameters of magnitude
$\approx \varepsilon$ [$M_A$ gauged by the generic electroweak scale].
Then, as long as neither $\kappa$, $\lambda$ nor the other scales
become too large, we observe a hierarchical structure in the CP--even
mass matrix of the form:
\be \{ M^2_{ij} \}= M_A^2 \left( 
\begin{array}{cc} 
A & \varepsilon C^{\dagger} \\ 
\varepsilon C  & \varepsilon^2 B
\end{array} \right), \ee
where $B$ is a $2 \times 2$ matrix, $C$ is a column vector and $A$ is
a scalar, all of order unity.

Performing an auxiliary unitary transformation defined by
the matrix,
\be V^{\dagger} = \left( 
\begin{array}{cc} 
1 - \frac{1}{2} \varepsilon^2 \Gamma^{\dagger} \Gamma 
& -\varepsilon \Gamma^{\dagger} \\ 
\varepsilon \Gamma & 1 \hspace{-0.16cm} 1 
- \frac{1}{2}\varepsilon^2 \Gamma \Gamma^{\dagger}
\end{array} \right)+ \mathcal{O} ( \varepsilon^4 ), \ee  
with $\Gamma = C/A,$ the mass matrix takes block diagonal form:
\ba VM^2V^{\dagger} &=& M_A^2 \left( 
\begin{array}{cc} 
A + \varepsilon^2 C^{\dagger}C/A  &0\\
0&\varepsilon^2 (B-CC^{\dagger}/A)
\end{array} \right) +\mathcal{O}(\varepsilon^3) \nonumber \\
&=&\left(
\begin{array}{ccc} 
M_{11}^2+\frac{M_{13}^4}{M_{11}^2} &0&0\\
0&M_{22}^2&M_{23}^2-\frac{M_{13}^2M_{12}^2}{M_{11}^2}\\
0&M_{23}^2-\frac{M_{13}^2M_{12}^2}{M_{11}^2}&
M_{33}^2-\frac{M_{13}^4}{M_{11}^2}
\end{array} \right) +\mathcal{O}(\varepsilon^3), \label{eq:vmv}
\ea
where $M^2_{ij}$ are the entries of the CP--even mass matrix,
Eqns.(\ref{eq:m11even}--\ref{eq:m33even})\footnote{In this section, we
are concerned only with the CP--even squared mass matrix and the
notation $M \equiv M_+$ is understood.}.  Note that $M^2_{12}/M_A^2$
is actually $\mathcal{O}(\varepsilon^3)$, so many terms can be
neglected (and have been in the above).

$VM^2V^{\dagger}$ is now easily diagonalized to give the approximate CP--even
Higgs boson masses:
\ba
M_{H_3}^2 &=& M_{11}^2+\frac{M_{13}^4}{M_{11}^2}, \label{eq:apph3} \\
M_{H_{2/1}}^2 &=& \half \left( M_{22}^2+M_{33}^2-\frac{M_{13}^4}{M_{11}^2}
\pm \sqrt{\left( M_{22}^2-M_{33}^2+\frac{M_{13}^4}{M_{11}^2} \right)^2
+4 \left(M_{23}^2-\frac{M_{13}^2M_{12}^2}{M_{11}^2} \right)^2} \right).
\label{eq:apph12} 
\ea
The diagonalization matrix is given by
\be
R= \left( 
\begin{array}{ccc} 
1 & 0 & 0 \\
0 & \cos \theta_H & -\sin \theta_H \\
0 & \sin \theta_H &  \cos \theta_H 
\end{array} \right) \hspace*{0.3cm} {\rm with} \hspace*{0.3cm}
\tan\theta_H=\frac{M_{23}^2
-\frac{M_{13}^2M_{12}^2}{M_{11}^2}}{M_{22}^2-M_{H_1}^2}.
\label{eq:tanthh} \ee
Combining $R$ with the first rotation matrix $V$ gives us the matrix
$O$ linking the states $S_i$ to the mass eigenstates (see Eqn.(\ref{eq:HtoS}): $H=OS$):
\ba
O&=&R^{\dagger}V \nonumber\\
&=&\left( 
\begin{array}{ccc} 
  1-M_{13}^4/2M_{11}^4 & M_{12}^2/M_{11}^2 & M_{13}^2/M_{11}^2 \\
 -\left[M_{13}^2\sin\theta_H+M_{12}^2\cos\theta_H\right]/M_{11}^2 &
 \phantom{-}\cos \theta_H &
 \sin\theta_H \left[ 1- M_{13}^4 /2M_{11}^4 \right] \\
 \phantom{-}\left[M_{12}^2\sin\theta_H-M_{13}^2\cos\theta_H\right]/M_{11}^2 &
 -\sin \theta_H &
 \cos\theta_H \left[ 1- M_{13}^4/2M_{11}^4 \right] \end{array} \right). \nonumber \\
&& \label{eq:Uapprox}
\ea

\begin{figure}[ht]
\bc \includegraphics[scale=0.6]{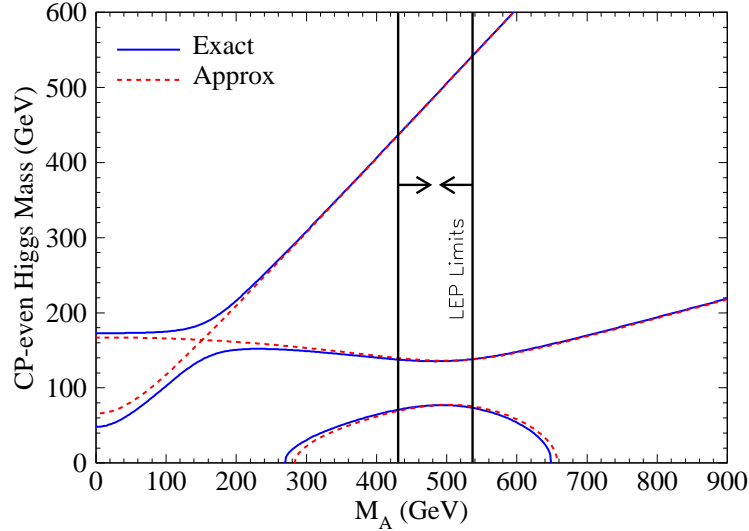} \ec 
\caption{\it The exact (numerical) solution and the approximate solution
         for the one-loop CP--even Higgs masses as a function of $M_A$ 
         for $\lambda=0.3$, $\kappa=0.1$, $v_s=3\,v$, $\tan \beta=3$ 
         and \mbox{$A_{\kappa}=-100$~GeV}.}
\label{fig:cpmapp}
\end{figure}

This approximation works remarkably well, as can be seen in
Fig.(\ref{fig:cpmapp}), where both the exact (numerical) solution and
the approximate solution, Eqns.(\ref{eq:apph3}--\ref{eq:apph12}), for
the one-loop CP--even Higgs masses are shown as a function of $M_A$
for the favoured parameter set of the slightly broken PQ
symmetry, $\lambda=0.3$, $\kappa=0.1$, $v_s=3\,v$, $\tan \beta=3$ and
$A_{\kappa}=-100$ GeV, introduced in Sec.(3.2). The exact and
approximate solutions agree well as long as $M_A$ does not become too
small.

\end{document}